\documentclass[preprint,12pt]{elsarticle}

\usepackage{lineno}
\usepackage{graphicx}
\usepackage{amssymb}
\usepackage{graphicx}
\usepackage[utf8]{inputenc}
\usepackage{lmodern}
\usepackage[T1]{fontenc}
\usepackage{lscape}
\usepackage{amsmath}	
\usepackage{amssymb}
\usepackage{cases}
\usepackage{graphicx}
\usepackage{caption}
\usepackage{comment}

\usepackage{makecell}
\usepackage{multirow}
\usepackage{amsfonts}
\usepackage{float} 
\usepackage{diagbox}
\usepackage{fullpage}
\usepackage{url}
\usepackage{rotating}
\usepackage{eurosym}
\usepackage{wrapfig}
\usepackage[final]{pdfpages} 
\usepackage{epstopdf} 
\usepackage[a4paper]{geometry}
\usepackage[nottoc]{tocbibind}
\usepackage{placeins}
\usepackage{float}
\usepackage{listings}
\usepackage{color, colortbl}
\usepackage{hyperref}
\usepackage{xcolor}
\usepackage{graphicx, caption, subcaption}
\usepackage{sidecap}
\usepackage[capitalise]{cleveref}
\usepackage{algorithm}
\usepackage{algpseudocode}
\usepackage{array, makecell} 
\usepackage{booktabs}
\usepackage{ulem}
\usepackage{bm}

\hypersetup{
colorlinks,
citecolor=black,
filecolor=black,
linkcolor=black,
urlcolor=black
}

\makeatletter
\renewcommand{\ALG@name}{Algorithm}
\makeatother

\biboptions{semicolon,authoryear}

\makeatletter
\def\ps@pprintTitle{%
  \let\@oddhead\@empty
  \let\@evenhead\@empty
  \def\@oddfoot{\reset@font\hfil\thepage\hfil} 
  \let\@evenfoot\@oddfoot
}
\makeatother

\abstracttitle{\centerline{Abstract}}
\begin{document}

\begin{frontmatter}


\title{A neural-network model-measurement-based observation operator for weather radar reflectivity assimilation}

\author{%
  \textbf{Marco Stefanelli}$^{1}$,
  \textbf{Žiga Zaplotnik}$^{2,1}$,
  \textbf{Gregor Skok}$^{1}$\\[0.5em]
  {\small
    $^{1}$University of Ljubljana, Faculty of Mathematics and Physics,\\
    Jadranska Cesta 19, 1000 Ljubljana, Slovenia\\
    $^{2}$European Centre for Medium-Range Weather Forecasts,\\
    Robert-Schuman-Platz 3, 53175 Bonn, Germany\\[0.4em]
    Corresponding author: Marco Stefanelli \\ \texttt{marco.stefanelli@fmf.uni-lj.si} 
  }%
}

\begin{abstract}
In three-dimensional variational data assimilation (3DVar) for numerical weather prediction (NWP), the observation operator $\mathcal{H}$ plays a central role by mapping model state variables to an observation equivalent field. For weather radar, however, specifying $\mathcal{H}$ is particularly challenging: reflectivity is a nonlinear, microphysics-dependent diagnostic quantity that only indirectly relates to the model’s prognostic variables, making traditional parameterised radar operators complex, regime-dependent and difficult to tune.

In this study, we propose a neural-network (NN)-based observation operator for radar reflectivity and apply it within a 3DVar data assimilation (DA) framework. Using five years (2019–2023) of radar reflectivity data from the Lisca radar and 4.4\,km-resolution short-range forecasts from ALADIN model over Slovenia, we train a convolutional encoder–decoder neural network to map model temperature, humidity, horizontal wind components and mean sea-level pressure fields to radar reflectivity. Statistical evaluation showed that the trained observation operator is able to represent the broader distribution and spatial occurrence of weak-to-moderate precipitation but remains limited in reproducing the frequency, intensity, and spatial organization of the strongest radar reflectivity features. 
The practical performance of the framework was evaluated through an extreme precipitation case, which caused widespread floods in Slovenia on August 4, 2023. In this case, assimilating the full radar disc reduces the domain-averaged reflectivity root-mean-square error (RMSE) from 5.99\,dB(Z) to 3.47\,dB(Z) and improves the alignment between the analysed and observed convective bands.

Embedded within 3DVar, the Jacobian of the NN observation operator allows radar reflectivity observations to inform model state variables, producing corresponding analysis increments. The proposed NN radar observation operator offers a flexible alternative to traditional parameterised radar operators for improving convective-storm forecasts.

\end{abstract}

\begin{keyword}
Data Assimilation \sep Observation Operator \sep Neural Network Observation Operator \sep Radar Data Assimilation

\end{keyword}

\end{frontmatter}

\newpage
\section{Introduction}
Recent years have witnessed growing interest in hybrid approaches that integrate machine learning (ML) techniques into traditional data assimilation (DA) frameworks, aiming to overcome structural and computational limitations in geophysical DA systems (e.g. see \citet{cheng2023machine} and \citep{pasmans2025ensemble} for a comprehensive review). ML methods have been successfully used to reformulate complex components of DA systems, such as background error covariance modelling (\citealp{melinc20243d, melinc2025unified}) or the observation operator for coupled ocean–acoustic variational data assimilation (\citealp{storto2021neural}). These developments suggest that ML can learn nonlinear relationships that are difficult to express analytically or to parameterise within classical statistical frameworks.

Among the various DA applications, radar data assimilation (DA) presents unique challenges. Despite the high spatial and temporal resolution of radar observations, their assimilation into numerical weather prediction (NWP) models remains problematic \citep{fabry2020radar}. Numerous variational and ensemble-based techniques have been developed to assimilate radar reflectivity and doppler velocity data (e.g., \citealp{stephan2008assimilation, dowell2011ensemble, wattrelot2014operational, liu2020direct, thiruvengadam2021radar}), yet the improvement in short-term convective forecasts remains limited.

As emphasised by \citet{fabry2020radar}, radar DA faces a fundamental representativeness problem. Precipitation fields exhibit intense small-scale variability and strong nonlinearity, which are only partially resolved by current NWP systems. This leads to systematic mismatches between observed and simulated radar reflectivity, especially at convective scales, where unresolved structures and imperfect microphysical parameterisations dominate errors \citep{fabry2020radar}. Consequently, simple extrapolation methods for radar nowcasting often outperform numerical forecasts with radar DA during the first few hours (FIG. 1 in \citealp{fabry2020radar}). These difficulties originate from a combination of model and system-related factors. On the modelling side, limited spatial resolution and simplified microphysical and convection schemes prevent the NWP model from representing the fine-scale variability captured by radar observations. On the DA side, the observation operator plays a critical role as the translator between model and observation spaces. When this operator, together with the underlying model physics, cannot reproduce radar reflectivity for the correct meteorological reasons, the innovations lose their physical meaning. As a result, the assimilation process may introduce corrections that do not accurately reflect the actual atmospheric state, thereby reducing the effectiveness of the analysis. Furthermore, in most current operational systems, the assimilation of radar reflectivity primarily corrects the hydrometeor or precipitation fields, with limited or no adjustment to the dynamical and thermodynamical variables (e.g., temperature, humidity, horizontal wind components). These corrections tend to decay rapidly during model integration, limiting their predictive value \citep{fabry2020radar}.

These limitations highlight the central role of the observation operator in determining the effectiveness of radar DA. Because it bridges the model and observation spaces, any structural inadequacy in the operator, such as oversimplified microphysics or nonlinear scattering relationships, directly constrains the physical consistency of the analysis. Conventional radar operators, which rely on parameterised descriptions of hydrometeor backscattering, are not only computationally demanding but also highly sensitive to assumptions about hydrometeor particle size distributions and scattering properties. As a result, they often struggle to reproduce observed reflectivities for the right physical reasons, reinforcing the representativeness problem described above.

These challenges underscore the need for a more flexible formulation of the radar observation operator. We propose a neural-network (NN)-based observation operator designed to learn the nonlinear mapping between dynamical and thermodynamic model state variables (temperature, humidity, wind components, and pressure) and radar reflectivity. This data-driven operator is embedded within a variational DA framework to improve the realism and efficiency of the forward process, ultimately enhancing the representativeness of innovations and the impact of radar DA on convective-scale forecasts.

The paper is organised as follows. Section~\ref{section:methodology} presents methodological aspects of our study, Section~\ref{section:results} presents the results. Discussion and Conclusions are given in Section~\ref{section:conclusions}.

\section{Methodology}\label{section:methodology}
 
In this study, we propose an NN–based observation operator for weather radar reflectivity and apply it in a 3DVar data assimilation.
The NN observation operator is trained to determine the expected reflectivity summed over the previous hour from short-range forecasts of temperature ($t$), horizontal wind components ($u$ and $v$), relative humidity ($r$), at four pressure levels (975\,hPa, 925\,hPa, 850\,hPa, and 800\,hPa) and three surface variables, 2\,m temperature ($t2m$), 2\,m relative humidity ($r2m$) and mean sea level pressure ($msl$) from the ALADIN numerical weather prediction model (hereafter 'ALADIN model outputs' refers to the listed fields) to the corresponding observed radar reflectivities. Once trained, the NN observation operator is used in 3DVar to produce model-equivalent of radar reflectivities and tested for independent cases not included in the training dataset. 

In this section, we first describe the radar and model datasets used to construct and evaluate the NN operator (\cref{{subsec:datasets}}: Dataset description). We then summarise the 3DVar formulation and its configuration for the present study (\cref{subsec:3DVar}: 3DVar formulation). The architecture and implementation of the NN-based observation operator are detailed in \cref{subsec:NNH}: The NN Observation Operator. Finally, the training strategy is presented in \cref{subsec:training}: Training setup.

\subsection{Dataset description}\label{subsec:datasets}

\begin{figure}[htb]
\centering

\begin{subfigure}[t]{0.48\textwidth}
    \centering
    \includegraphics[height=5.2cm,keepaspectratio]{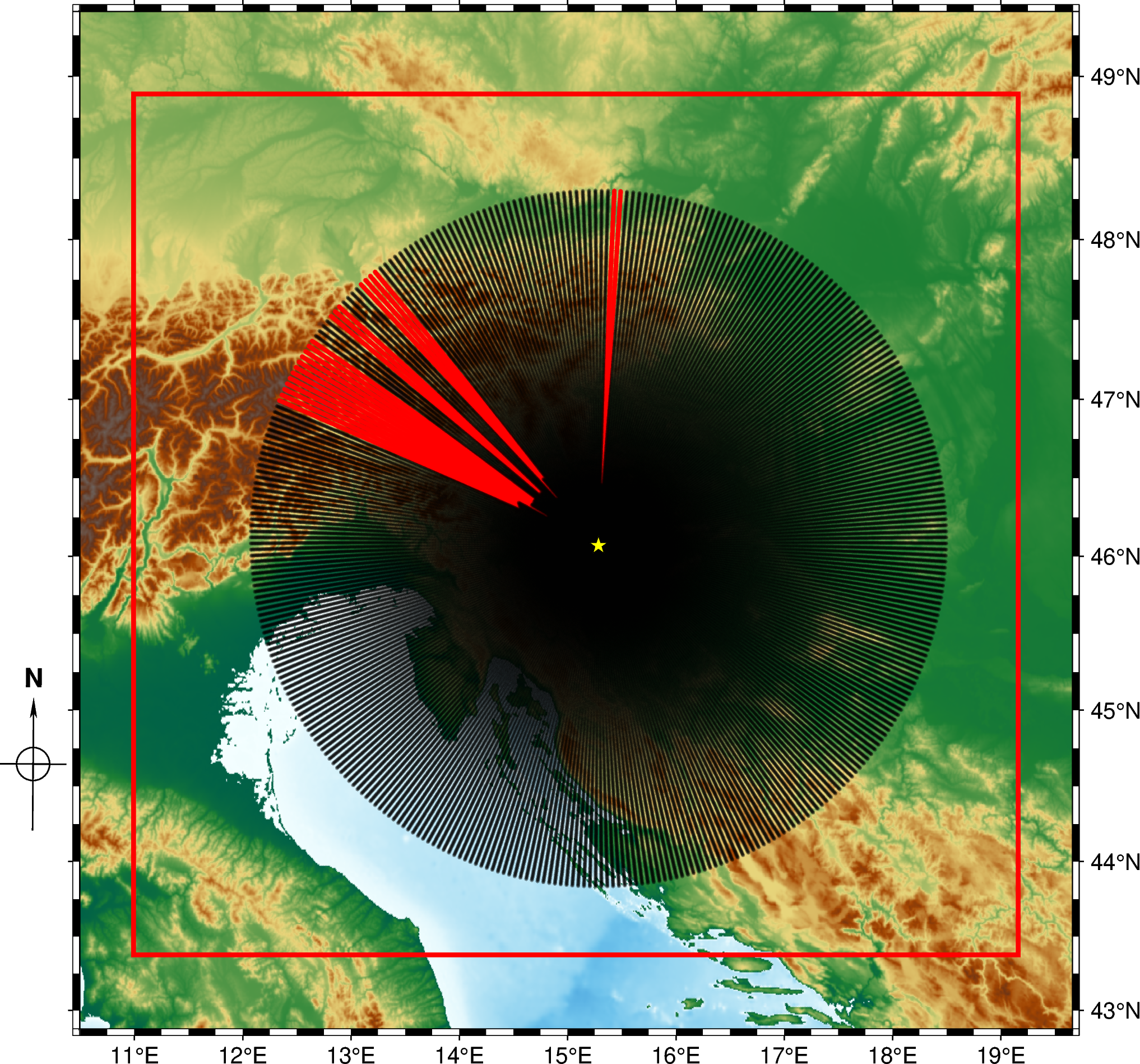}
    \caption{}
    \label{fig:domains}
\end{subfigure}
\hfill
\begin{subfigure}[t]{0.48\textwidth}
    \centering
    \raisebox{1.0cm}{%
        \includegraphics[height=2.7cm,keepaspectratio]{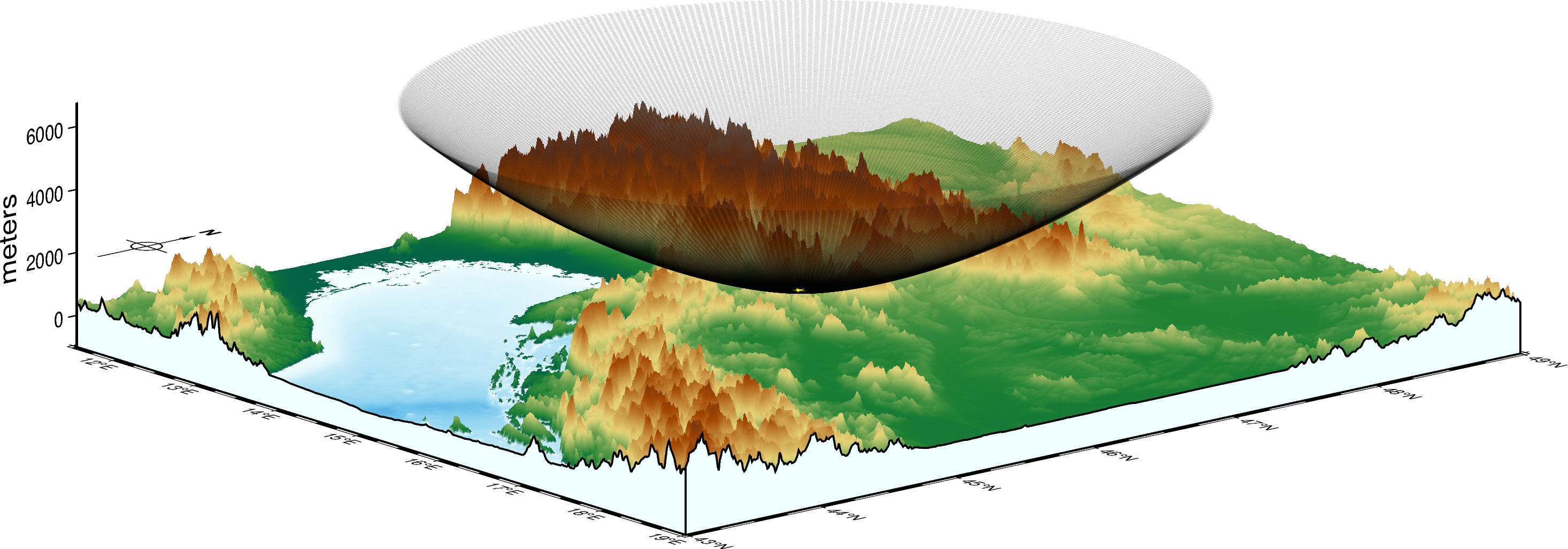}%
    }
    \caption{}
    \label{fig:radar}
\end{subfigure}

\caption{Spatial overview of the datasets used for training the neural-network observation operator. (a) spatial domain and radar sampling. The red square indicates the spatial extent of the ALADIN data used in the analysis, and the black dots indicate the coverage of the Lisca radar of $0.5^\circ$ elevation. Red areas on the radar disc indicate regions where terrain mountain shadow zones block the radar beam. (b) 3D geometry of the Lisca radar’s first elevation scan ($0.5^\circ$).}
\label{fig:datasets}
\end{figure}

To develop the proposed NN observation operator, we used five years (2019–2023) of ALADIN model outputs and weather radar data from the Lisca radar (Slovenia, 46.07$^\circ$N, 15.29$^\circ$E, 945\,m a.s.l., frequency band: C, \cite{Strinar2018}; further details at \citealp{wmo-radar}), both provided by the Slovenian Met Office (ARSO). The ALADIN model \citep{Termonia2018} data were available at a horizontal resolution of 4.4\,km and a temporal resolution of 1\,h and 6\,h lead time, with a 6\,h analysis cycle (00, 06, 12, 18 UTC). From the radar dataset, we selected the lowest elevation angle ($0.5^\circ$) and computed the 3-D coordinates (x, y, z) using the Python package \textit{xradar} \citep{grover_2025_15862221} which refers to \citet{doviak2014doppler} for mathematical formulation. Both the ALADIN and radar domains are shown in \cref{fig:datasets}. Red areas on the radar disc indicate regions where the radar beam is blocked by terrain, i.e. the mountain shadow zones. These occur mostly to the northwest in the lee of the Kamnik-Savinja Alps.

Before their use in model training, radar reflectivity data were preprocessed to ensure temporal and spatial consistency, as well as overall data quality. Following the operational quality-control procedures used at Slovenian Met Office, two filters were applied: (1) only observations within a 160\,km horizontal radius of the radar site were retained, which corresponds to roughly the first 3\,km of the atmospheric column, and (2) reflectivity values below 13.5\,dBZ were discarded and set to NaN to suppress background noise and spurious echoes. To ensure consistency between ALADIN fields and the observed atmospheric volume, we therefore restricted the input model variables to the lower troposphere. While upper-level winds and thermodynamic perturbations also affect deep convection, their consistent inclusion would require observations from additional radar elevation angles and the corresponding higher model levels.

\begin{figure}[htb]
\centering
    \begin{subfigure}[h]{0.4\textwidth}
        \includegraphics[width=\textwidth]{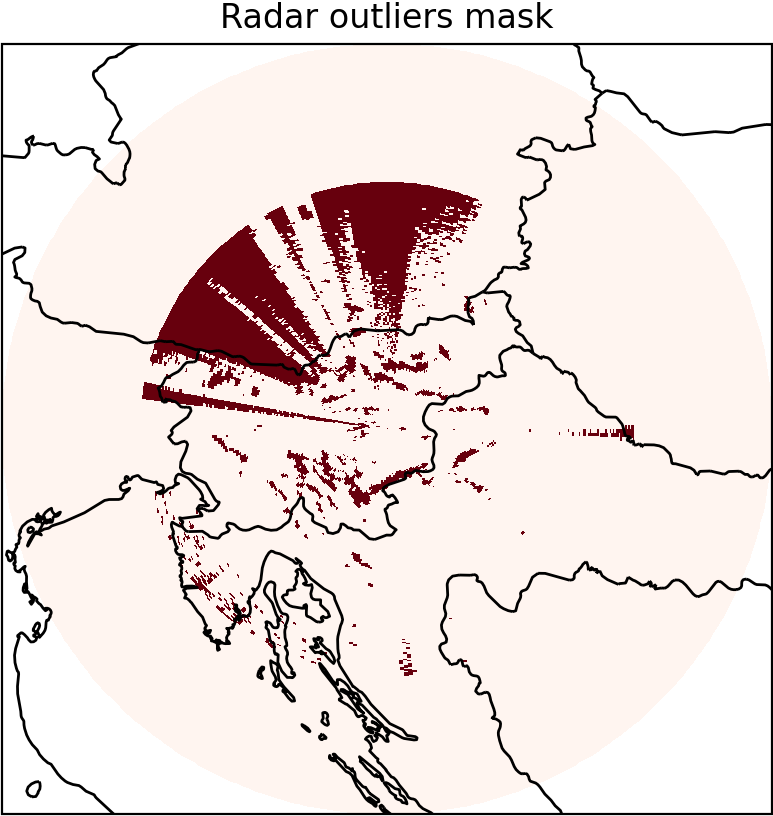}
    \end{subfigure}
\caption{Spatial distribution of the global radar outlier mask derived from Mahalanobis-distance analysis of reflectivity data. The mask was computed only for data points within a 160\,km radius of the Lisca radar that had already passed the initial filtering steps. Gates exceeding the 90th percentile distance threshold were classified as outliers (dark red) and excluded from subsequent use to enhance data quality.}
\label{fig:mask}
\end{figure}

To further enhance data quality, a global outlier mask was constructed using a histogram-based Mahalanobis distance approach as in \citet{franch2020taasrad19}. For each radar resolution gate (i.e. each fixed position in the azimuth–range radar grid), we constructed a normalised histogram of reflectivity using 526 bins of width 0.1\,dB(Z), based on a random subset comprising 20\% of all available time steps. Each gate’s histogram was then treated as a feature vector in an $N_\text{bins}$-dimensional space. Using all gates, we estimated the global mean histogram and covariance matrix, and subsequently computed the Mahalanobis distance for each gate $i$ as follows:
\begin{equation}
D_i = \sqrt{(\mathbf{x}_i - \boldsymbol{\mu})^{\mathrm{T}} \boldsymbol{\Sigma}^{-1} (\mathbf{x}_i - \boldsymbol{\mu})},
\label{eq:mahalanobis}
\end{equation}
In \cref{eq:mahalanobis}, $\mathbf{x}_i$ denotes the histogram vector at gate $i$, $\boldsymbol{\mu}$ is the global mean histogram vector, and $\boldsymbol{\Sigma}$ is the covariance matrix of all gates. To ensure numerical stability, a small regularisation term ($\varepsilon = 10^{-6}$) was added to the diagonal of $\boldsymbol{\Sigma}$ prior to inversion. Gates with distances exceeding the 90th percentile threshold (i.e., the upper 10\% of the distribution) were classified as outliers. The resulting mask (\cref{fig:mask}) was then applied to remove these outlier gates from the dataset. A strong spatial correspondence is observed between the outlier mask derived from the Mahalanobis distance analysis (\cref{fig:mask}) and the terrain-induced radar beam blockage areas (Fig.\ref{fig:datasets}-a). This correlation suggests that a significant fraction of the identified outliers originates from regions affected by partial or complete signal obstruction by mountainous terrain, highlighting the method’s sensitivity to detecting physically meaningful artefacts in the radar observations.

\begin{figure}[htb]
\centering
    \begin{subfigure}[h]{0.99\textwidth}
        \includegraphics[width=\textwidth]{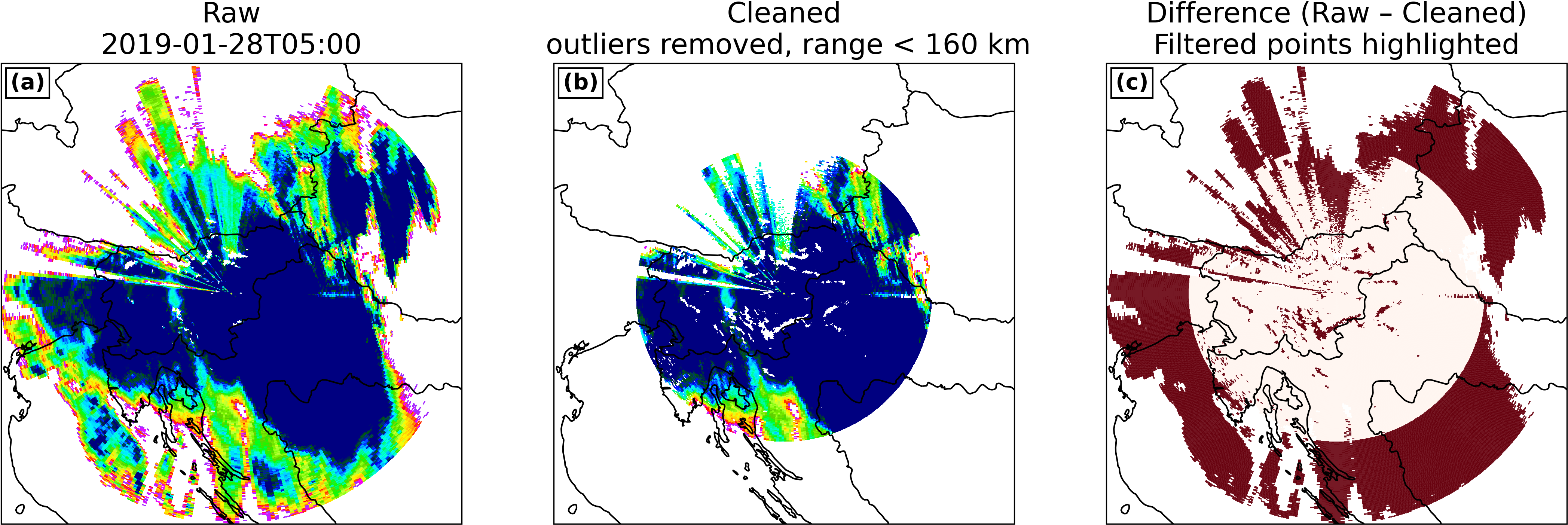}
    \end{subfigure}
\caption{Example of the radar–data cleaning procedure. (a) raw reflectivity field showing radial artefacts, distant-range noise, and terrain-induced beam blockage effects. (b) cleaned field after applying the radial filtering, retaining only the observations in a disc of 160\,km from the radar centre, and the Mahalanobis-distance outlier mask. (c) difference between the raw and cleaned fields, with filtered gates highlighted in dark red.}
\label{fig:cleaned_datasets}
\end{figure}

The practical impact of the cleaning procedures is illustrated in \cref{fig:cleaned_datasets}, which shows a representative example of the raw radar field (\cref{fig:cleaned_datasets}-a), the cleaned dataset (\cref{fig:cleaned_datasets}-b), and the corresponding difference map (\cref{fig:cleaned_datasets}-c), highlighting the filtered points. The raw data exhibit numerous radial streaks and azimuth–range sectors contaminated by non-meteorological echoes, many of which coincide with regions of known beam blockage. After applying the radial filter and the Mahalanobis-based outlier mask, these artefacts are effectively removed, yielding a spatially coherent precipitation field with substantially reduced noise. The difference panel highlights the removed gates in red tones.

\begin{figure}[htb]
\centering
    \begin{subfigure}[h]{0.5\textwidth}
        \includegraphics[width=\textwidth]{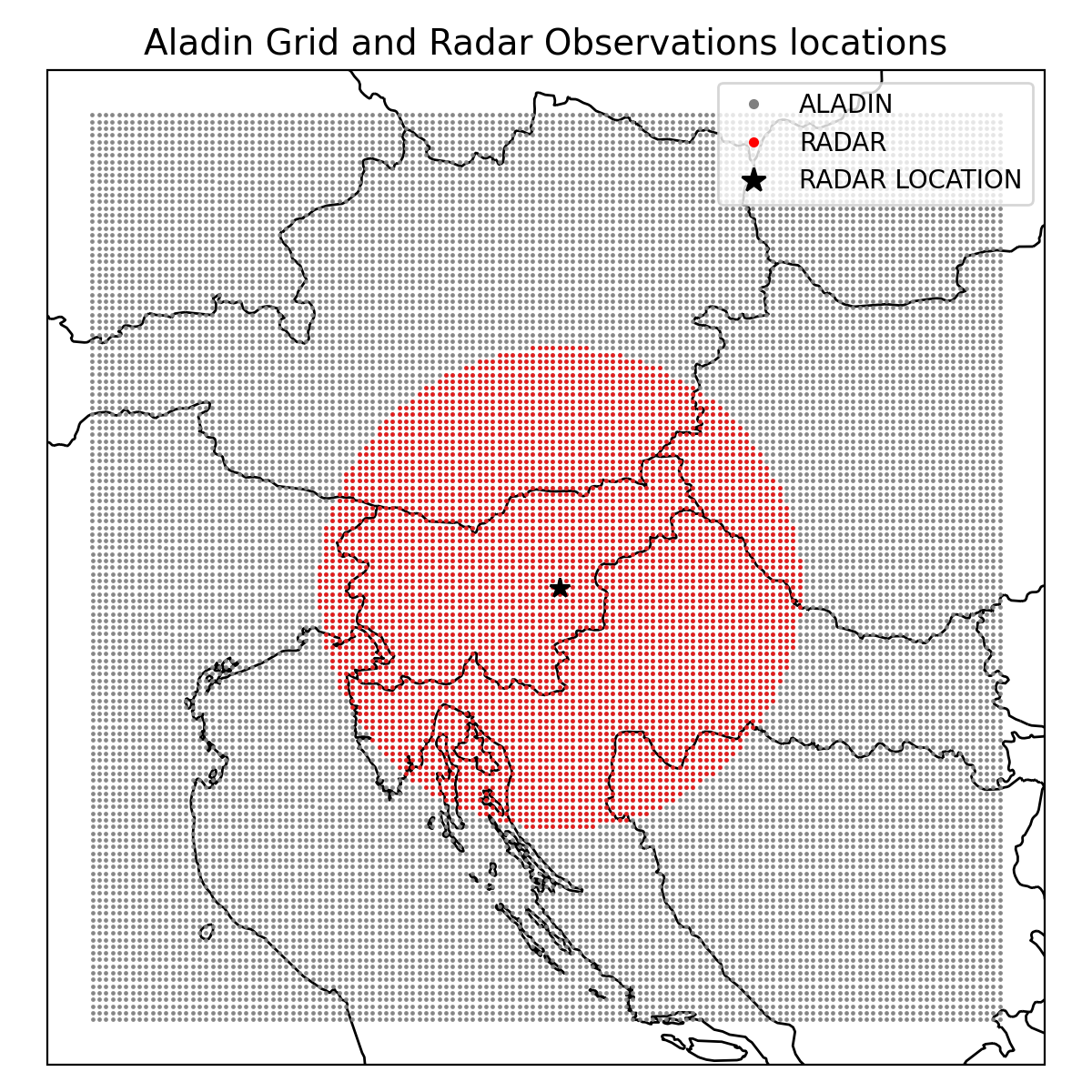}
    \end{subfigure}
\caption{Spatial correspondence between the ALADIN model and radar datasets: black dots mark the model grid, and red dots denote radar observations interpolated onto the same spatial domain.}
\label{fig:grid}
\end{figure}

The quality-controlled radar observations were subsequently summed in linear units (Z [$mm^6\,m^{-3}$]) over each hour to obtain a temporally more representative target at the ALADIN model output frequency; the resulting field was then reported again in reflectivity units. This procedure has been preferred instead of using the radar image closest to the analysis time for several considerations. First, ALADIN outputs precipitation at time t as 1-hour accumulation
of precipitation between time t and (t - 1\,hr). Our approach would allow a qualitative comparison between radar reflectivity simulated from model precipitation (e.g. through the use of a Z–R relation) and radar reflectivity simulated with our NN observation operator from outputs of basic model variables at time t. Second, hourly aggregation yields smoother reflectivity fields, which are generally easier for a neural network to learn than noisy fields with
sharp gradients. Finally, the observed reflectivity field exhibits larger spatio-temporal variability than the model output (the effective temporal resolution of 4.4 km ALADIN is
approximately 15 mins due to numerical diffusion), so we assume that the instantaneous model state at time t is representative of a broader range of precipitation observations within the corresponding hourly window. The reflectivity sum was then spatially interpolated onto the model grid (by taking the value from the nearest radar point) to ensure spatial consistency between the two datasets. Finally, since radar observations have some missing timesteps, to maintain temporal coherence, the ALADIN dataset was filtered to retain only those timesteps for which corresponding radar observations were available.  The final datasets grid is drawn in \cref{fig:grid}. Both datasets are available on Zenodo \citep{Zenododatasets}.

\subsection{3DVar formulation}\label{subsec:3DVar}

In the 3DVar framework \citep{lorenc1986analysis}, the analysis state is obtained by minimising a cost function $\mathcal{J}(\mathbf{x})$ that measures the distance of the model state $\mathbf{x}$ to the background state $\mathbf{x}_{\mathrm{b}}$ and observations $\mathbf{y}$:
\begin{equation}
\mathcal{J}(\mathbf{x}) =
\frac{1}{2}(\mathbf{x}-\mathbf{x}_{\mathrm{b}})^{\mathrm{T}}\mathbf{B}^{-1}(\mathbf{x}-\mathbf{x}_{\mathrm{b}}) + \frac{1}{2}(\mathbf{y}-\mathcal{H}(\mathbf{x}))^{\mathrm{T}}\mathbf{R}^{-1}(\mathbf{y}-\mathcal{H}(\mathbf{x}))
\label{eq:J}
\end{equation}
Here, $\mathbf{B}$ and $\mathbf{R}$ denote the background and observation error covariance matrices, respectively, while $\mathcal{H}$ is the non-linear observation operator that projects the model state into the observation space. 

For observations that directly correspond to model variables, such as radiosonde-measured temperature or horizontal wind vector, $\mathcal{H}$ typically consists only of interpolation and coordinate transformation procedures \citep{lahoz2010data}.
However, when the observed quantity does not directly correspond to a model variable, as in the case of radar reflectivity, the construction of $\mathcal{H}$ becomes considerably more complex. This is the case of the ALADIN 3DVar scheme for radar reflectivity data \citep{wattrelot2014operational}, which retrieves pseudo-observations of relative humidity from observed vertical profiles of reflectivity using a one-dimensional Bayesian inversion. The resulting humidity pseudo-observations are subsequently assimilated in the AROME 3DVar system \citep{brousseau2011background}.

Minimising \cref{eq:J} requires computing its gradient, which vanishes at the minimum of the cost function:
\begin{equation}
\nabla \mathbf{\mathcal{J}(x)} = \mathbf{B}^{-1}(\mathbf{x}-\mathbf{x}_{\mathrm{b}}) - \left( \frac{\partial \mathbf{\mathcal{H}(x)}}{\partial \mathbf{x}} \right)^T \mathbf{R^{-1}} (\mathbf{y} - \mathbf{\mathcal{H}(x)}) \, .
\label{eq:gJ}
\end{equation}
The Jacobian of $\mathcal{H}$, i.e. $\mathbf{H} = \partial \mathcal{H}(\mathbf{x})/\partial\mathbf{x}$, governs how observational information is backpropagated to the variables used to construct the observation operator during the minimisation.

For radar reflectivity, an accurate and differentiable observation operator is therefore essential to ensure physically consistent increments and to enhance the quality of 
analyses and forecasts.

In practice, the minimisation of \cref{eq:J} is performed in the incremental form \citep{courtier1997variational} and \cref{eq:J} reads as follows:

\begin{equation}
\mathcal{J}(\mathbf{\bm{\chi}})
= \frac{1}{2}\,\bm{\chi}^\top \bm{\chi}
\;+\; \frac{1}{2}\,\big(\mathbf{y}-\mathcal{H}(\mathbf{x}_{\mathrm{b}}+\mathbf{L}\bm{\chi})\big)^\top
\mathbf{R}^{-1} \, 
\big(\mathbf{y}-\mathcal{H}(\mathbf{x}_{\mathrm{b}}+\mathbf{L}\bm{\chi})\big) \,  
\label{eq:Jinc}
\end{equation}
where $\bm{\chi}$ is the control vector and $\mathbf{L}$ the control variable transform (CVT) operator, such that 
\begin{equation}
\delta\mathbf{x} = \mathbf{x} - \mathbf{x}_{\mathrm{b}} = \mathbf{L}\,\bm{\chi}
\label{eq:inc}
\end{equation}
In the defined control space, $\mathbf{B}$ can be factorised as follows:
\begin{equation}
\mathbf{B} = \mathbf{L}\,\mathbf{L}^{\mathrm{T}}
\label{eq:B_factorization}
\end{equation}
In our implementation, we model the control operator $\mathbf{L}$ as
\begin{equation}
\mathbf{L} = \mathbf{S}\,\mathbf{R\!F}
\label{eq:L_factorization}
\end{equation}
where $\mathbf{S}$ is a diagonal matrix containing the background-error variances of each model variable at each grid point and level, and $\mathbf{R\!F}$ is the first-order recursive filter (RF) operator \citep{,lorenc1992iterative, hayden1995recursive, purser2003numerical} which defines the auto-covariance. 

$\mathbf{S}$ is estimated from differences between forecasts and corresponding analyses valid at the same time, following a standard forecast–analysis statistics approach.
For each season, we consider one representative month of ALADIN model output in 2019 (January, April, July and October for winter, spring, summer and autumn, respectively). For each month, 6\,h lead-time forecast fields are paired with the corresponding analysis fields that are valid at the same synoptic times (00, 06, 12 and 18~UTC).
Finally, for each variable $q$ ($t$, $u$, $v$, $r$, $t2m$, $r2m$, $msl$) and the 4 pressure levels (975~hPa, 925~hPa, 850~hPa and 800~hPa) of 3-D variables, we compute the variance: 

\begin{equation}
\mathbf{S} =
\mathrm{Var}\bigl[q^{f} - q^{a}\bigr]
\label{eq:B_diag}
\end{equation}
where the superscripts $f$ and $a$ indicate forecast and analysis.

The RF is a simple iterative algorithm that requires only a few iterations to approximate the Gaussian function. The algorithm itself is 1-dimensional, but it can be straightforwardly applied in two or more dimensions by alternating dimensions between iterations \citep{purser2003numerical}. The RF methodology is 
used in oceanographic \citep{dobricic2008oceanographic, storto2021neural, stefanelli2025data} and atmospheric \citep{barker2004three, descombes2015generalized} 3DVar applications.

We intentionally adopt this simple, univariate, flow-independent model of $\mathbf{B}$, 
with an isotropic RF in the horizontal, in order to keep the 3DVar configuration lightweight and robust as a proof-of-concept test case for the NN-based observation operator. All 3DVar experiments presented in the \cref{section:results} use this specification of the control space and background-error covariance matrix with 4 RF iterations and a correlation radius of 10\,km. We assumed the vertical autocovariances and cross-covariances to be zero.

In our experiments, the observation error covariance matrix $\mathbf{R}$ is assumed diagonal (the observation errors are not correlated) with observation error standard deviation of 2\,dB(Z).

\subsection{The NN-based observation operator}\label{subsec:NNH}

The nonlinear observation operator, \(\mathcal{H}\), is implemented as a convolutional encoder–decoder NN that models the mapping from ALADIN model subspace to the radar reflectivity observation space. The rationale behind this approach is to utilise the most accurrate estimates of the true atmospheric state for both the input and target output. For the input, we use analyses as they represent the most comprehensive and physically-consistent estimate of the atmosphere. No single observing system provides a comparable description of the instantaneous, full atmospheric state without data voids. 
For the target output, we intentionally select observed radar reflectivities as the ground-truth representation of the precipitation field, rather than relying on simulated reflectivities from a traditional forward model.

\begin{figure}[htb]
\centering
    \begin{subfigure}[h]{0.99\textwidth}
        \includegraphics[width=\textwidth]{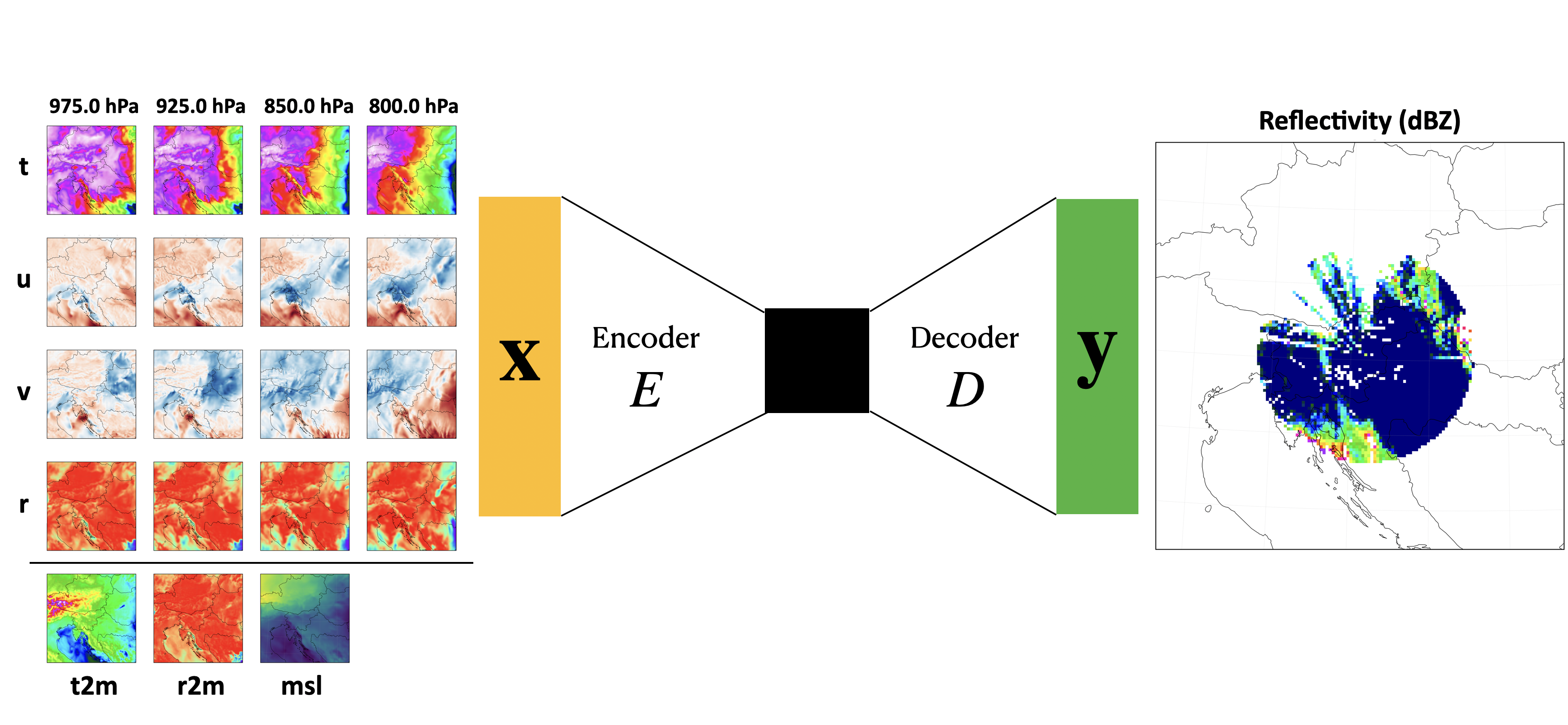}
    \end{subfigure}
\caption{Schematic overview of the convolutional encoder–decoder neural network used to emulate the 3DVar observation operator. The model is trained with input $\mathbf{x}$ consisting of ALADIN temperature ($t$), relative humidity ($r$), and horizontal wind components ($u$ and $v$) at four pressure levels (975\,hPa, 925\,hPa, 850\,hPa, and 800\,hPa), together with 2\,m temperature ($t2m$), 2\,m relative humidity ($r2m$), and mean sea-level pressure ($msl$). The output ${\mathbf{y}}$ is the reflectivity field measured by the radar, interpolated onto the model grid at the corresponding time.}
\label{fig:resunet}
\end{figure}

Formally, the operator approximates the transformation
\begin{equation}\label{eq:obs_operator}
\hat{\mathbf{y}} = D\bigl(E(\mathbf{x})\bigr)=\mathcal{H}(\mathbf{x}) \, ,
\end{equation}
where $D$ denotes the decoder and $E$ the encoder. $\mathbf{x}$ is the input vector of the neural network, which consists of the following ALADIN model fields: temperature ($t$), horizontal wind components ($u$) and ($v$) and relative humidity ($r$) at the first four pressure levels (975\,hPa, 925\,hPa, 850\,hPa, and 800\,hPa). In addition, 2\,m temperature ($t2m$), 2\,m relative humidity ($r2m$), and mean sea-level pressure ($msl$) are provided to the network. These variables are chosen to capture the thermodynamic and dynamic conditions relevant to the precipitation formation. After the encoding-decoding process, the model-equivalent of the reflectivity field, $\hat{\mathbf{y}}$, is obtained. A schematic overview of the architecture is shown in \cref{fig:resunet}.

The NN architecture is implemented as a residual U-Net (ResUNet, \citealp{he2015deep}). 
The ResUNet follows an encoder–decoder design with skip connections, residual blocks, and Squeeze-and-Excitation units \citep{hu2018squeeze} that apply channel-wise attention. 

Training is guided by an edge-aware, spatially weighted loss. A Gaussian radial weight emphasises regions close to the radar, where observations are more reliable, while a boundary-enhancement map increases the loss along precipitation edges, where sharp reflectivity gradients occur. These weights modulate a composite loss combining Huber, gradient-difference, and Structural Similarity Index Measure (SSIM) terms, encouraging accurate amplitudes, sharp boundaries and structurally coherent patterns. 
For a detailed explanation of the ResUNet architecture and the implemented loss function, the reader is referred to \ref{appendices:A}.

\subsection{Training Setup}\label{subsec:training}

To train the neural-network observation operator, we use analysis times (00, 06, 12, and 18 UTC) from 2019 to 2022 (5706 timesteps in total) as the best estimate of the atmospheric state, shuffled to avoid spurious correlations between consecutive timesteps. These timesteps were split as follows: 4564 samples for training, 571 for validation, and 571 for an initial test set used for a rapid check after training different models and model tuning.  All the timesteps of 2023 (8725 in total) are held out as an independent test set for statistical evaluation and 3DVar experiments on selected events.

To increase the representation of precipitation, and particularly of intense convective events, in the training dataset, we applied targeted data augmentation \citep{aggarwal2018neural}. The augmenting procedure assigns each training sample between one and three augmentation cycles according to both the maximum reflectivity within the domain and the spatial extent of moderate-to-strong echoes. The latter is quantified by the number of grid points exceeding selected reflectivity thresholds. The severity-dependent augmentation criteria are summarised in \ref{crit:augmentation_criteria}.

\begin{equation}
\mathrm{Cycles} =
\left\{
\begin{aligned}
1\ \rightarrow \ & \mathrm{max}(\mathrm{Reflectivity})\geq25~\mathrm{dBZ}
\ \land\ \mathrm{pixels}(\mathrm{Reflectivity}\geq25~\mathrm{dBZ})\geq5,\\
1\ \rightarrow \ & \mathrm{max}(\mathrm{Reflectivity})\geq35~\mathrm{dBZ}
\ \land\ \mathrm{pixels}(\mathrm{Reflectivity}\geq25~\mathrm{dBZ})\geq20,\\
2\ \rightarrow \ & \mathrm{max}(\mathrm{Reflectivity})\geq45~\mathrm{dBZ}
\ \land\ \mathrm{pixels}(\mathrm{Reflectivity}\geq35~\mathrm{dBZ})\geq10,\\
3\ \rightarrow \ & \mathrm{max}(\mathrm{Reflectivity})\geq55~\mathrm{dBZ}
\ \land\ \mathrm{pixels}(\mathrm{Reflectivity}\geq45~\mathrm{dBZ})\geq5,
\end{aligned}
\right.
\label{crit:augmentation_criteria}
\end{equation}

Samples that do not satisfy any of these criteria are not augmented. In this way, increasingly intense and spatially extensive reflectivity events receive progressively greater representation in the training set.

Each augmentation cycle contains four orientations, corresponding to spatial rotations of \(0^\circ\), \(90^\circ\), \(180^\circ\), and \(270^\circ\) counter-clockwise about the vertical axis. The \(0^\circ\) orientation therefore corresponds to a duplicate of the original spatial fields. All scalar meteorological predictors and the corresponding radar-reflectivity target are rotated identically, preserving their spatial correspondence. Wind components are instead treated explicitly as vector pairs. Given the orientation of the model grid, with the \(x\)-axis increasing eastward and the \(y\)-axis increasing northward, every spatial rotation is accompanied by the corresponding physical transformation of each $u$-$v$ pair as in \cref{eq:wind_rotation}.
\begin{equation}
\left\{
\begin{aligned}
u' &= -v, \quad &&v'=u,  \quad &&\text{for }90^\circ,\\
u' &= -u, \quad &&v'=-v, \quad &&\text{for }180^\circ,\\
u' &=  v, \quad &&v'=-u, \quad &&\text{for }270^\circ.
\end{aligned}
\right.
\label{eq:wind_rotation}
\end{equation}
Thus, the augmented samples preserve both predictor--target spatial consistency and the physical orientation of the horizontal wind vectors.

The first augmentation cycle consists exclusively of these exact physical rotations of the meteorological input fields and reflectivity target. For samples assigned a second or third cycle, additional weak perturbations are applied to the meteorological predictors to generate distinct realisations of the same atmospheric situation. These perturbations are generated independently for each augmentation cycle and each sample. For every scalar input channel ($X$), we first compute the spatial standard deviation, \(\sigma_X\), over the corresponding \(137\times137\) field. A \(137\times137\) Gaussian random field is then generated and smoothed four successive times using a local \(3\times3\) averaging filter to suppress grid-scale fluctuations and introduce spatial coherence. The resulting field is subsequently normalised to zero mean and unit standard deviation. The perturbed scalar field is defined in \cref{eq:scalar_pert}.
\begin{equation}
X^{*}(i,j)
=
X_{\mathrm{rot}}(i,j)
+
0.005\,\sigma_X\,\eta(i,j),
\label{eq:scalar_pert}
\end{equation}
where $i$ and $j$ are the grid indices, \(X_{\mathrm{rot}}(i,j)\) denotes the spatially rotated scalar field, \(\sigma_X\) is its sample-specific spatial standard deviation, and \(\eta(i,j)\) is the spatially smooth random perturbation normalized to zero mean and unit standard deviation. 

For wind predictors, the perturbation amplitude is determined from the variability of the wind speed rather than from $u$ and $v$ independently. For each $u$-$v$ pair, the wind-speed field is calculated, and its spatial standard deviation, \(\sigma_V\), is evaluated over the individual sample. Two independent spatially smoothed and normalised random fields, \(\eta_u(i,j)\) and \(\eta_v(i,j)\), are then generated for the two wind components. After the physical vector rotation described above, the perturbed components are given by
\begin{equation}
u^{*}(i,j)
=
u_{\mathrm{rot}}(i,j)
+
0.01\,\sigma_V\,\eta_u(i,j),
\end{equation}
and
\begin{equation}
v^{*}(i,j)
=
v_{\mathrm{rot}}(i,j)
+
0.01\,\sigma_V\,\eta_v(i,j),
\end{equation}
where \(u_{\mathrm{rot}}\) and \(v_{\mathrm{rot}}\) are the physically rotated wind components. 

The coefficients \(0.005\) and \(0.01\) are two hyperparameters and control the amplitude of the perturbations relative to the intrinsic spatial variability of each sample. They correspond to perturbation amplitudes of \(0.5\%\) of the sample-specific spatial standard deviation for scalar variables and \(1\%\) of the spatial standard deviation of wind speed for the wind components, respectively. The larger factor used for the wind components compensates for the fact that the perturbation scale is derived from the wind variability and then applied separately to \(u\) and \(v\). These small values were chosen to introduce variability among otherwise repeated augmented samples while preserving the large-scale and mesoscale structure of the original meteorological state. The use of weak additive noise as a data-augmentation strategy has previously been adopted in meteorological deep-learning applications; for example, \cite{lagerquist2020deep} applied Gaussian perturbations to radar predictors during CNN training, with the perturbation amplitude treated as an augmentation hyperparameter. In the present implementation, the perturbations are additionally scaled by the variability of each individual sample and spatially smoothed to avoid introducing unrealistic grid-scale noise.

The reflectivity target is not perturbed during the second and third augmentation cycles. These additional cycles are intended to represent small variations of the meteorological input state associated with the same observed radar situation; perturbing the reflectivity field itself would instead introduce an artificial modification of the observation. The corresponding target is therefore only spatially rotated, while weak stochastic perturbations are restricted to the input predictors. This approach also encourages the network to generalise the mapping between inputs and targets, analogous to the regularising effect of latent-space perturbations applied during training.

Applying these severity-dependent criteria, 2287 of the 4564 original training samples were selected for augmentation, while 2277 samples remained unaugmented. Among the selected samples, 1340 were assigned one augmentation cycle, 663 two cycles, and 284 three cycles. The first cycle generated \(2287\times4=9148\) exact rotated samples. The additional cycles applied to the more intense cases generated a further $(663 + 2\times284)\times4 = 4924$ perturbed samples. The revised augmentation procedure therefore adds 14\,072 samples to the original 4564-member training dataset, resulting in a final training set of 18\,636 samples. Thus, the corresponding input tensor has dimensions \((18\,636,137,137,19)\), and the reflectivity target has dimensions \((18\,636,137,137)\). The validation and test datasets are not augmented, ensuring that model evaluation is performed exclusively on independent, unmodified samples.

The network is trained with a batch size of 16 using the Adam optimiser and an initial learning rate of $10^{-3}$. Training proceeds for a maximum of 1000 epochs, with early stopping applied if the validation loss does not improve for 10 consecutive epochs. In addition, an adaptive learning rate scheduler is employed. If the validation loss does not decrease for 3 epochs, the learning rate is reduced by a factor of 0.1 to improve convergence in the later stages of training (see \ref{appendices:A} for further details). The model is implemented in PyTorch, and the code is available on Zenodo \citep{Zenodo3DHNNmodel}.

\section{Results}\label{section:results}

The NN observation operator for radar reflectivities was first qualitatively assessed for different precipitation regimes (\cref{subsec:performances}). A quantitative statistic evaluation is shown in \cref{subsec:stat}. To evaluate its performance in a data assimilation framework, we next examine its behaviour in a range of data assimilation settings. In \cref{subsec:blob_obs}, we analyse how assimilated radar observations induce analysis increments in the model variables through the gradient of the observation operator. For this purpose, we select a small subdomain of radar volumes (consisting of 250 observation points) to characterise the spatial structure, amplitude, and multivariate properties of the resulting increments.
Finally, \cref{subsec:slo_case} presents a Slovenian flood case study that documents the system's performance during an extreme precipitation event and assesses whether the use of $\mathcal{H}$ produces dynamically consistent analyses that better capture intense convective structures relevant to high-impact flooding. 
All the codes and $\mathbf{B}$ matrices used are available on Zenodo \citep{Zenodo3DVar}.

\subsection{Performance in different precipitation regimes}\label{subsec:performances}

\begin{figure}[htb]
\centering
    \begin{subfigure}[h]{\textwidth}
        \includegraphics[width=\textwidth]{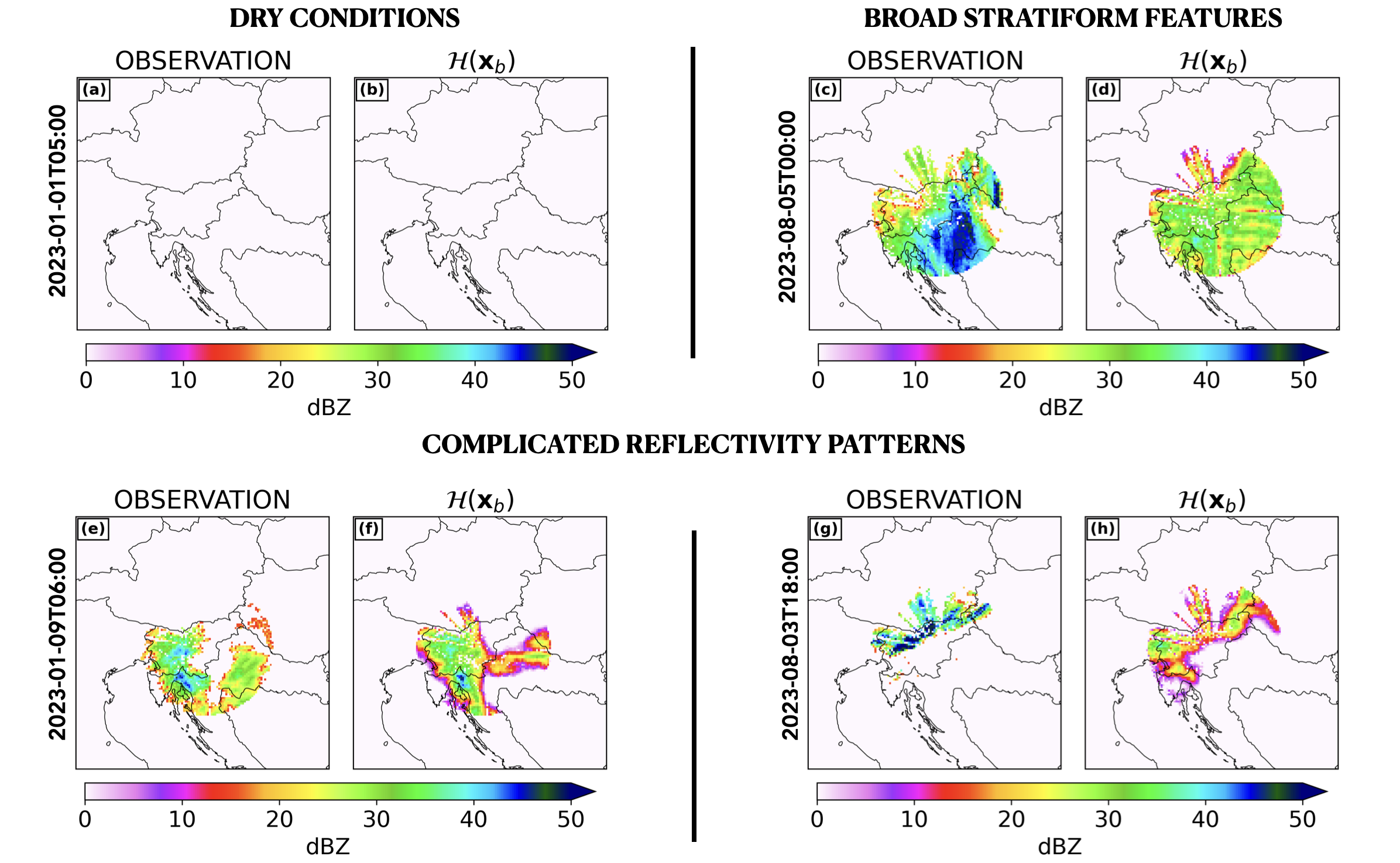}
    \end{subfigure}
\caption{Generated versus observed radar reflectivity for four test samples. a, c, e, g show the radar reflectivity used as target observation and b, d, f, h the corresponding generated reflectivity field obtained from the background state, \(\mathcal{H}(\mathbf{x}_b)\). a-b: dry conditions (no radar echo) on 2023-01-01T05:00. c-d: broad stratiform precipitation on 2023-08-05T00:00. e-h: two cases with complicated reflectivity patterns, including embedded convective structures on 2023-01-09T06:00 and 2023-08-03T18:00.}
\label{fig:HNN_res}
\end{figure}

To qualitatively assess how the NN-based observation operator behaves under different precipitation conditions, we inspected three cases with different characteristic precipitation regimes over the year 2023 (not used during the training of the operator): dry conditions (\cref{fig:HNN_res}~a,b); broad stratiform precipitation (\cref{fig:HNN_res}~c,d); complicated reflectivity patterns (\cref{fig:HNN_res}~e-h).
For each case, the observed radar reflectivity (used as target) is shown alongside the model-equivalent reflectivity field. The latter is obtained by applying the NN-based observation operator to the background state, $\mathcal{H}(\mathbf{x}_b)$.

Under dry conditions (2023-01-01T05:00, \cref{fig:HNN_res}~a,b), both the observed reflectivity and its model equivalent $\mathcal{H}(\mathbf{x}_b)$ are free of echoes. The NN operator correctly reproduces the absence of precipitation, indicating that it does not introduce spurious reflectivity under clear-sky conditions.

For broad stratiform precipitation (2023-08-05T00:00, \cref{fig:HNN_res}~c,d), the operator captures the overall spatial extent of the radar echo and the large-scale reflectivity gradients. The main regions of moderate reflectivity are reasonably reproduced, although small-scale details differ between the two fields.

In cases with complicated reflectivity patterns (2023-01-09T06:00 and 2023-08-03T18:00, \cref{fig:HNN_res}~e-h), associated with embedded convective elements and narrow bands, the network still reproduces the general spatial organisation of the echo, including the location and orientation of the main features. Nevertheless, differences appear in the fine-scale structure and local intensity, with some displacement and smoothing of individual convective cores. These examples illustrate that the NN-based operator performs consistently across regimes, from no-precipitation to complex convective situations.

\subsection{Statistical evaluation} \label{subsec:stat}

The performance of the trained NN-based observation operator was evaluated by comparing its reflectivity fields with the corresponding radar observations over the radar-disc domain in 2023. The objective of this analysis was to assess not only whether the operator reproduces the overall occurrence of precipitation echoes, but also how well it captures the intensity-dependent frequency, intensity-defined events, and spatial consistency of reflectivity patterns. To this end, we considered three complementary diagnostics: (i) cumulative coverage as a function of reflectivity threshold, (ii) Categorical Verification Scores \citep[Probability Of Detection (POD), Critical Success Index (CSI) and False Alarm Ratio (FAR), ][]{Wilks2019}, and (iii) the Fractions Skill Score \citep[FSS,][]{Roberts2008,Roberts2008a}, a widely used neighborhood-based spatial verification metric.

\begin{figure}[htb]
\centering
    \begin{subfigure}[h]{\textwidth}
        \includegraphics[width=\textwidth]{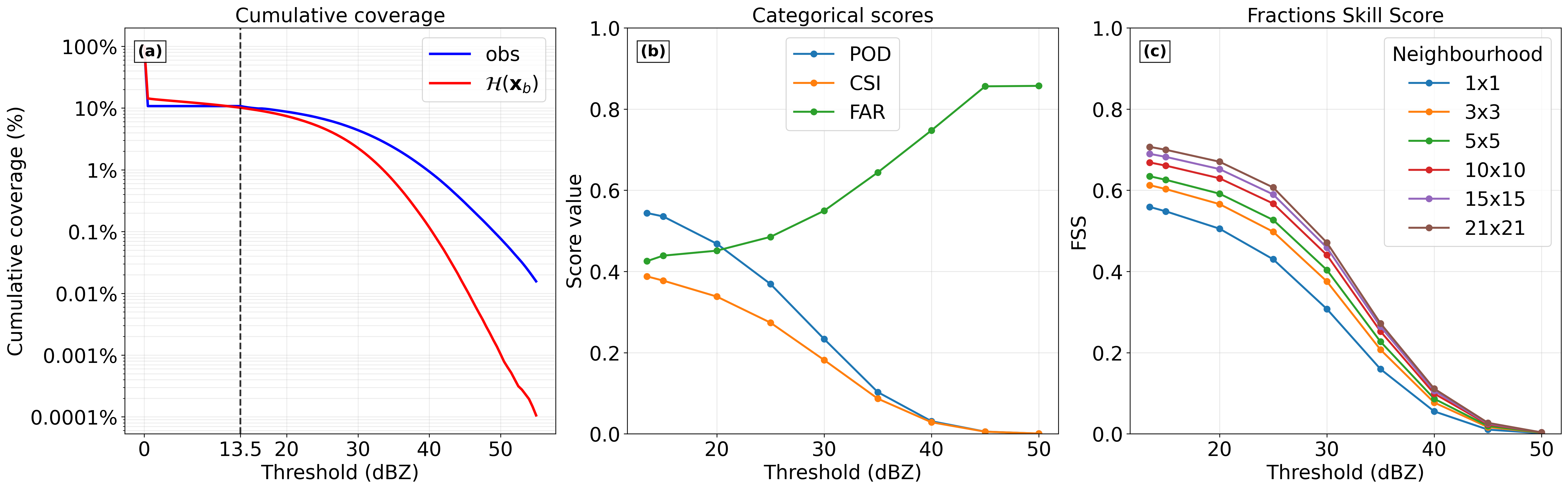}
    \end{subfigure}
\caption{Statistical evaluation of the NN-based observation operator $\mathcal{H}(\mathbf{x}_b)$ against radar observations over the radar-disc domain in 2023. (a) Cumulative reflectivity coverage, defined as the fraction of all valid radar-disc pixels (in all timesteps) with reflectivity greater than or equal to a given threshold ([0, 55] dBZ with a step of 0.5 dBZ), shown for the observations (blue) and $\mathcal{H}(\mathbf{x}_b)$ (red). The dashed vertical line marks the 13.5~dBZ threshold used to distinguish precipitation-related echoes from dry or cloud-only conditions. (b) Pixel-based categorical verification scores, namely the probability of detection (POD), critical success index (CSI), and false alarm ratio (FAR), computed for thresholds ranging from 13.5 to 50~dBZ. (c) Fractions Skill Score (FSS) as a function of threshold for square-shaped neighbourhoods ranging from $1\times1$ to $21\times21$ grid points. For panels (b) and (c), statistics were computed only for timesteps satisfying the conditional selection criterion, i.e. timesteps for which either the observed or modeled reflectivity exceeded a minimum areal coverage of 1\% of the radar-disc pixels at the corresponding reflectivity threshold.}
\label{fig:obsop_stats}
\end{figure}

Figure~\ref{fig:obsop_stats}a shows the cumulative coverage of reflectivity for the observations (blue) and $\mathcal{H}(\mathbf{x}_b)$ (red), computed over all radar disc pixels for all timesteps in 2023 shown as a function of reflectivity threshold. The thresholds span the interval [0, 55] dBZ with a step of 0.5 dBZ. The coverage at a certain threshold is defined as the fraction of pixels (in all timesteps) whose reflectivity is equal or larger than the threshold. Because the metric is cumulative, the curve decreases monotonically with increasing threshold and summarizes the exceedance frequency of progressively stronger echoes. The logarithmic y-axis enhances the visibility of differences in the low-coverage regime and in the upper-reflectivity tail. The dashed vertical line indicates the 13.5 dBZ threshold which is the minimum non-zero reflectivity (used to separate precipiation events from dry and cloud-only events). The figure indicates that the observations and $\mathcal{H}(\mathbf{x}_b)$ have similar cumulative coverage at low reflectivity thresholds, with both curves close to 10\% up to 20 dBZ. For larger thresholds, the coverage of $\mathcal{H}(\mathbf{x}_b)$ reduces more rapidly than the observations, revealing a deficit of moderate and intense echoes. The discrepancy becomes marked above 30~dBZ and is strongest in the high-reflectivity tail, where the logarithmic scale emphasizes the substantially lower exceedance frequency in $\mathcal{H}(\mathbf{x}_b)$. This suggests that the observation operator reflectivity captures the lower-to-medium reflectivity coverage reasonably well but underestimates the occurrence of stronger radar echoes.

To complement this distribution-based view with event-based verification, we computed categorical verification scores as a function of reflectivity threshold for NN-based observation operator $\mathcal{H}(\mathbf{x}_b)$ relative to observations computed over the radar disc domain considering all the available timesteps and for thresholds of 13.5, 15, 20, 25, 30, 35, 40, 45, and 50 dBZ in 2023 (Fig.~\ref{fig:obsop_stats}b). For each threshold and at every timestep, the script identifies the number of pixels exceeding the threshold in both observations and model space. A timestep-selection filter is applied before computing the average, a timestep is kept only if the observations or the model reach a minimum echo coverage of 1\% of the radar disc. This isolates timesteps with meaningful precipitation echoes. The computed scores show a clear degradation of model skill as the reflectivity threshold increases. POD decreases steadily from about 0.54 at 13.5 dBZ to nearly zero above 40 dBZ, indicating that the model progressively misses a larger fraction of observed high-reflectivity events. CSI follows the same behavior, dropping from about 0.38 at low threshold to almost zero at the highest thresholds, which reflects an overall loss of agreement of intense events in the model and observations. In contrast, FAR increases monotonically with threshold, from roughly 0.44 at 13.5 dBZ to about 0.95 at 50 dBZ, showing that most modeled exceedances at high thresholds are not matched by observations. Together, these results indicate that the model retains moderate skill for weak-to-moderate reflectivity but performs poorly for strong convective cores, where events become both harder to detect and increasingly dominated by false alarms.

Since categorical scores only compare values at collocated locations, they are highly sensitive to spatial displacement errors \citep[e.g.,  the so-called 'double penalty' issue, ][]{Brown2011,Skok2022} and we additionally evaluated the spatial correspondence between observed and modeled reflectivity using the FSS (Fig.~\ref{fig:obsop_stats}c). For this purpose, the observed and modeled fields were thresholded using the same set of reflectivity levels used in the categorical analysis, and neighbourhood event fractions were computed using square-shaped neighbourhoods of sizes 1$\times$1, 3$\times$3, 5$\times$5, 10$\times$10, 15$\times$15, and 21$\times$21 grid points. Same as for the categorical scores, only timesteps satisfying the minimum 1\% areal-coverage criterion in either observations or model space were retained, so the FSS results also correspond to a conditioned sample of precipitation-relevant cases.

The FSS results confirm the threshold dependence already seen. At the lowest thresholds (13.5--20~dBZ), the FSS values are moderate and increase with neighbourhood size, indicating that part of the disagreement between the observations and $\mathcal{H}(\mathbf{x}_b)$ can be attributed to small spatial displacements rather than a complete mismatch or absence of precipitation signal. The results indicate that the trained observation operator can capture the broader spatial organization of weak and moderate echoes, even if the exact location of these features is not always perfectly reproduced at the grid-point level. However, the FSS decreases steadily with increasing threshold for all neighbourhood sizes. Around 30~dBZ, the score already drops substantially, indicating reduced skill in reproducing the position and extent of stronger precipitation patterns. At thresholds above 40~dBZ, the FSS becomes very small regardless of neighbourhood size, demonstrating that the most intense echoes are only rarely matched even when substantial spatial tolerance is allowed. This behaviour shows that the deficiencies of $\mathcal{H}(\mathbf{x}_b)$ at high reflectivity are not merely due to small displacement errors, but also reflect an inability of producing of intense events in general, which is in line with the analysis of the cumulative coverage shown in Figure~\ref{fig:obsop_stats}a.

Taken together, the three diagnostics provide a consistent picture of the strengths and limitations of the trained observation operator. The cumulative coverage analysis shows that $\mathcal{H}(\mathbf{x}_b)$ reproduces the overall frequency of weak-to-moderate reflectivity reasonably well but underestimates the occurrence of strong echoes. The categorical scores demonstrate that event detection skill decreases rapidly with increasing threshold, with both declining detection capability and increasing false alarms for intense events. The FSS further indicates that some of the mismatches at low thresholds can be explained by small-scale spatial displacements, whereas the lack of skill at high thresholds persists even for large neighbourhoods. Overall, these results show that the trained observation operator is able to represent the broader distribution and spatial occurrence of weak-to-moderate precipitation, but it remains limited in its ability to reproduce the frequency, intensity, and spatial organization of the strongest radar reflectivity features. This outcome is broadly consistent with the intended role of the trained observation operator, whose objective is not to reproduce the observed reflectivity exactly, but to provide a consistent mapping between model space and observation space while retaining the main statistical and spatial characteristics of the observations. From this perspective, the verification results are in line with expectations: they show that the operator reproduces the general reflectivity behaviour satisfactorily, while larger errors remain for the most intense and spatially localized events.

\subsection{Increments in model state variables, induced by $\mathcal{H}$}\label{subsec:blob_obs}

\begin{figure}
    \centering

    \begin{subfigure}[h]{0.99\textwidth}
        \includegraphics[width=\textwidth]{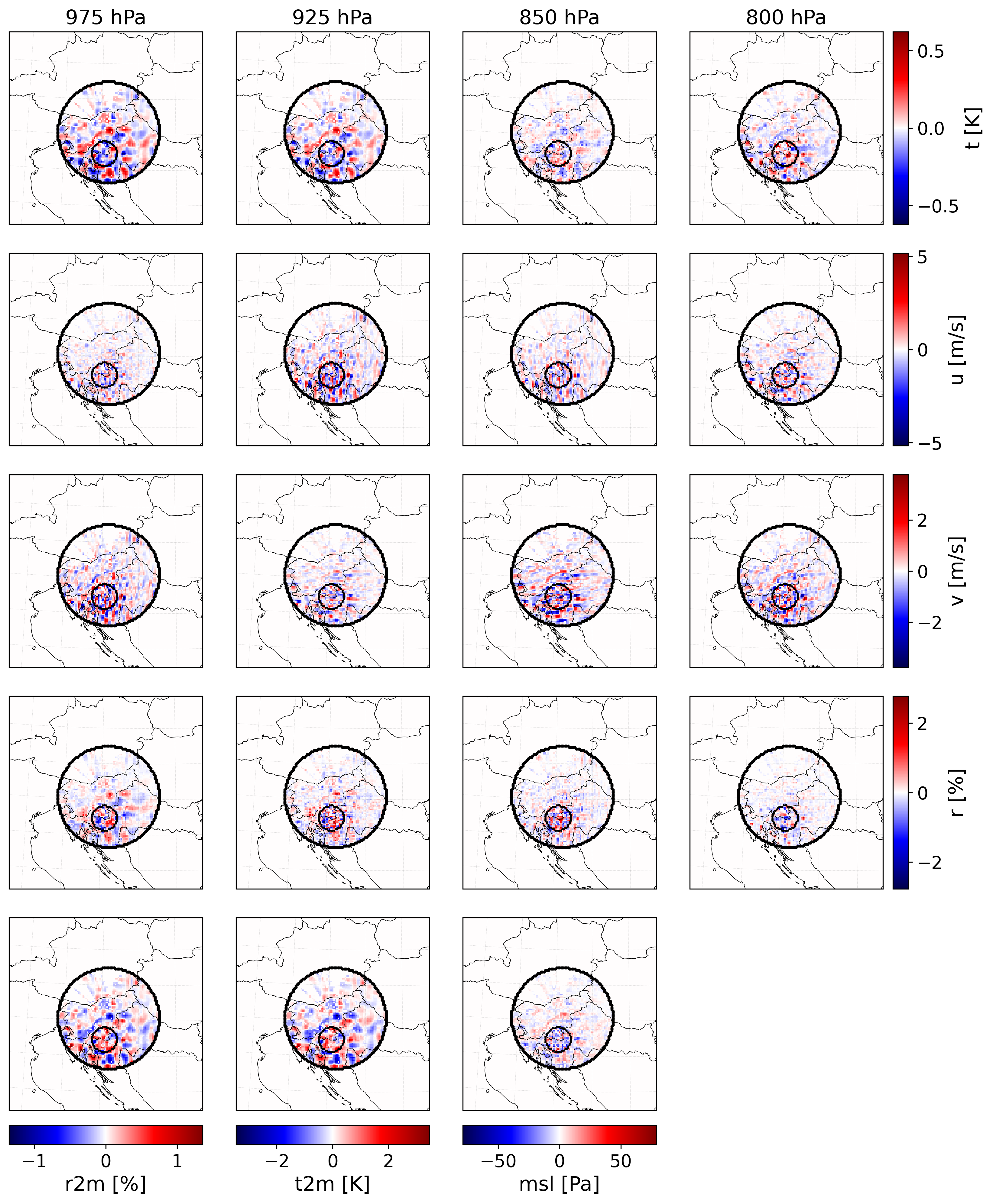}
    \end{subfigure}
    \caption{Analysis increments generated by the assimilation of a compact cluster of 250 reflectivity pixels and setting $\mathbf{B}=\mathbf{I}$. Columns show four pressure levels (975\,hPa, 925\,hPa, 850\,hPa, and 800\,hPa), while rows correspond to temperature $t$, zonal wind $u$, meridional wind $v$, relative humidity $r$, 2\,m relative humidity ($r2m$), 2\,m temperature ($t2m$), and mean sea-level pressure ($msl$). The outer black circle denotes the radar disc and the inner circle the location of the assimilated cluster.}
    \label{fig:increments_blob_250}
\end{figure}

To further investigate how the convolutional architecture of the NN-based observation operator spreads information in model space, we perform an idealised experiment in which a compact cluster of radar reflectivities is assimilated. The cluster consists of $N=250$ contiguous pixels (\cref{fig:increments_blob_250}, small black disc).

To isolate the effect of $\mathcal{H}$ itself, the background-error covariance matrix is set to the identity ($\mathbf{B} = \mathbf{I}$), such that no spatial spreading arises from the RF component of $\mathbf{B}$-matrix. Because the  observation operator is defined to produce model-equivalent reflectivity values over the entire radar disc domain (see Eq.~\ref{eq:obs_operator}), whereas $\mathbf{y}$ represents only a subset of that domain, the cost function (\ref{eq:J}) is reformulated as
\begin{equation}
    \mathcal{J}(\mathbf{x})=\frac{1}{2}(\mathbf{x}-\mathbf{x}_{\mathrm{b}})^{\mathrm{T}}(\mathbf{x}-\mathbf{x}_{\mathrm{b}}) + \frac{1}{2}(\mathbf{y}-\mathbf{M}\mathcal{H}(\mathbf{x}))^{\mathrm{T}}\,\mathbf{R}^{-1}
    \,(\mathbf{y}-\mathbf{M}\mathcal{H}(\mathbf{x})) \, ,
\end{equation}
where $\mathbf{M}$ is a masking matrix consisting of $N$ row vectors, each containing zeros and a single element with a value of 1 representing the observed points. A localisation mask is also defined within $\nabla\mathcal{J}(\mathbf{x})$. It is equal to 1 at all grid points inside the radar disc and 0 elsewhere (\cref{fig:increments_blob_250}, large black disc). This mask ensures that only model variables at grid points within the disc can be updated during the cost function minimisation, while those outside remain fixed.

\begin{figure}
    \centering

    \begin{subfigure}[h]{0.99\textwidth}
        \includegraphics[width=\textwidth]{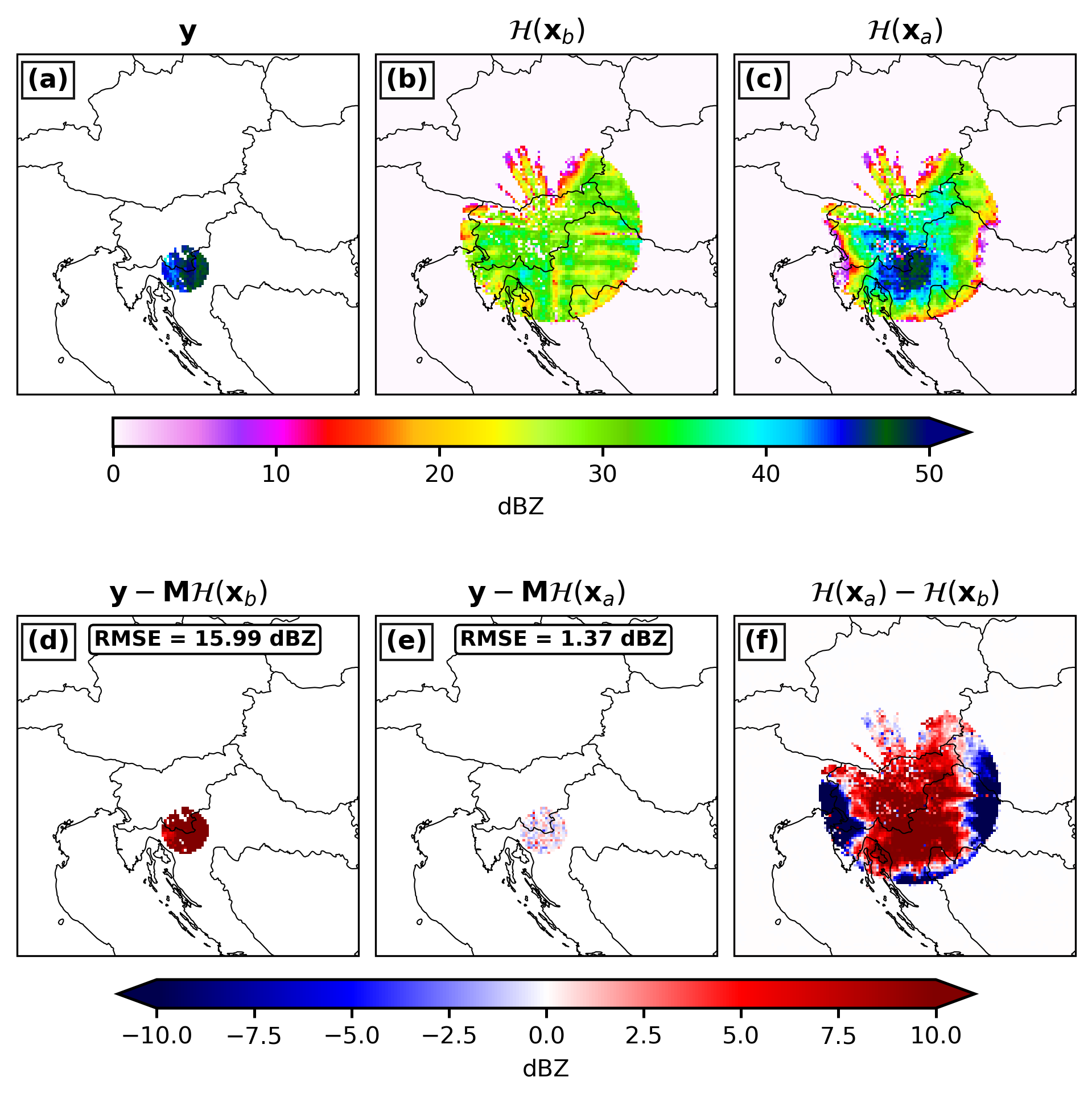}
    \end{subfigure}
     \caption{Evaluation of the NN-based observation operator for the experiment with a compact cluster of 250 assimilated pixels and $\mathbf{B}=\mathbf{I}$. Top row: (a) observed radar reflectivity, (b) background model-equivalent reflectivity $\mathcal{H}(\mathbf{x}_b)$, and (c) analysis model-equivalent reflectivity $\mathcal{H}(\mathbf{x}_a)$. Bottom row: (d) innovation $\mathrm{\mathbf{y}} - \mathbf{M}\mathcal{H}(\mathbf{x}_b)$. In the region of the observations location, RMSE $=15.99$\,dB(Z), (e) residual $\mathrm{\mathbf{y}} - \mathbf{M}\mathcal{H}(\mathbf{x}_a)$ with RMSE $=1.37$\,dB(Z) in the region of the observations location, and (f) change in model-equivalent reflectivity $\mathcal{H}(\mathbf{x}_a)-\mathcal{H}(\mathbf{x}_b)$.}
    \label{fig:Hxb_Hxa_blob_masked_250}
\end{figure}

\cref{fig:increments_blob_250} displays the resulting analysis increments and \cref{fig:Hxb_Hxa_blob_masked_250}-d the innovation, $\mathrm{\mathbf{y}} - \mathbf{M}\mathcal{H}(\mathbf{x}_b)$, that produced them. The increments form a compact yet clearly multivariate structure aligned with the observation cluster and largely confined to the radar disc. 
Some fields, most notably the near-surface temperature and relative humidity, exhibit weaker but non-negligible increments outside the immediate cluster, reflecting the influence of the CNN architecture of the trained $\mathcal{H}$ operator.

Because in this experiment $\mathbf{B} = \mathbf{I}$, the increments pattern are entirely generated by the convolutional structure of $\mathcal{H}$. The results, therefore, demonstrate that, despite its CNN architecture, the NN-based observation operator produces increments that remain mainly localised around the assimilated reflectivity cluster, with limited extensions into the surrounding area, rather than inducing spurious long-range adjustments.

To assess how these model-space increments translate back into observation space, \cref{fig:Hxb_Hxa_blob_masked_250} compares the reflectivity observations with the background and analysis equivalents from the 250-pixel cluster experiment. The top row shows the observed reflectivity (\cref{fig:Hxb_Hxa_blob_masked_250}-a), the model equivalent reflectivity from the model background state, $\mathcal{H}(\mathbf{x}_b)$ (\cref{fig:Hxb_Hxa_blob_masked_250}-b), and the model equivalent reflectivity from the analysis state,
$\mathcal{H}(\mathbf{x}_a)$ (\cref{fig:Hxb_Hxa_blob_masked_250}-c). The background markedly underestimates the observed echo, producing a smoother and weaker pattern over the cluster disc. After assimilation (with $\mathbf{B} = \mathbf{I}$), the analysis equivalent $\mathcal{H}(\mathbf{x}_a)$ closely reproduces the observed reflectivity field, including the location and intensity of the main high-reflectivity cores.

The bottom row highlights the corresponding differences. \cref{fig:Hxb_Hxa_blob_masked_250}-d shows the innovation which is predominantly positive and large over the cluster disc, with a root-mean-square error (RMSE) of the observed region of $15.99$\,dB(Z), reflecting the strong background underestimation. After assimilation, the residual in the observed region, $\mathrm{\mathbf{y}} - \mathbf{M}\mathcal{H}(\mathbf{x}_a)$ (\cref{fig:Hxb_Hxa_blob_masked_250}-e)  is reduced with an RMSE of $1.37$\,dB(Z). The change in model-equivalent reflectivity over the full radar disc,
$\mathcal{H}(\mathbf{x}_a)-\mathcal{H}(\mathbf{x}_b)$ is shown in \cref{fig:Hxb_Hxa_blob_masked_250}-f. Together with the increment patterns in model space, this demonstrates that $\mathcal{H}$ maps the local reflectivity cluster into a consistent multivariate response. 

It is important to stress that, because $\mathbf{B} = \mathbf{I}$ and no balance constraints are imposed, the resulting increments are not dynamically or thermodynamically consistent and should not be interpreted as a realistic atmospheric state. This experiment is purely diagnostic: it isolates the impact of the NN-based observation operator on the analysis and reveals its localisation and mapping properties, without any claim of physical realism in the resulting fields.

\subsection{Slovenian floods case study}\label{subsec:slo_case}

To illustrate the behaviour of the NN-based observation operator in a realistic high-impact weather event, we consider a case study on 4 August 2023 at 00UTC, characterised by widespread, intense precipitation over the radar domain that led to severe flooding and damage in Slovenia (\citealp{ARSO2023_august_floods}). During this event, the 500\,hPa geopotential field featured a warm ridge over southern Europe with an embedded trough over the Alps. This provided large-scale dynamical forcing with W-SW advection of moisture from the anomalously warm Mediterranean. The Dinaric-Alpine orography further enhanced convective development and rainfall intensity. The precipitation pattern was associated with a quasi-stationary triggering of convective cells (see \cref{fig:HNN_res}g, showing observed reflectivity on 3 August 2023 at 18 UTC, i.e. 6\,h before the time of our case-study). 

\begin{figure}
    \centering

    \begin{subfigure}[h]{0.99\textwidth}
        \includegraphics[width=\textwidth]{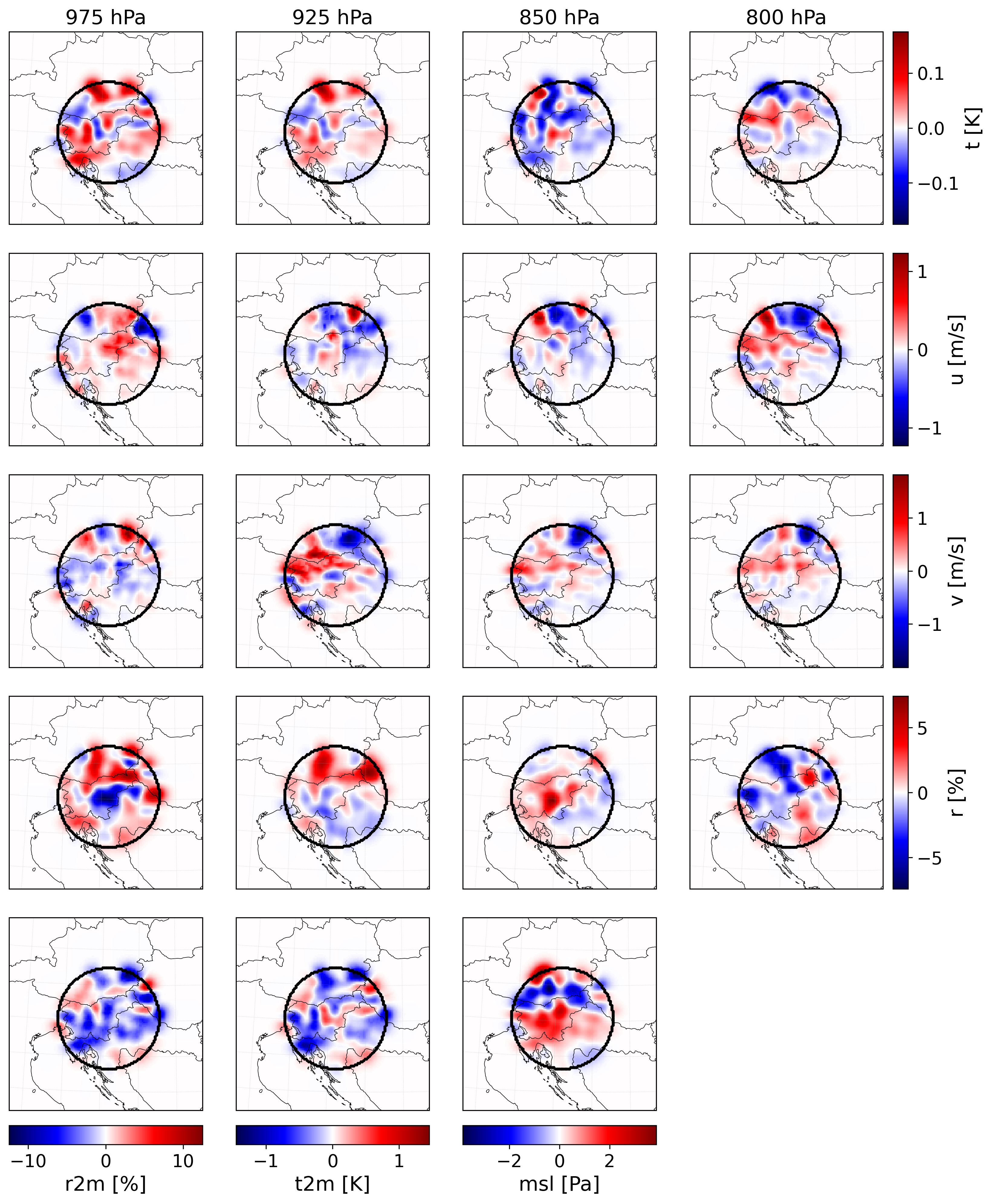}
    \end{subfigure}
    \caption{3DVar analysis increments for the Slovenian floods case on 4~August~2023 at 00~UTC. Panels show increments of temperature ($t$), meridional wind ($v$), zonal wind ($u$) and relative humidity ($r$) at four pressure levels (975~hPa, 925~hPa, 850~hPa and 800~hPa), together with increments of 2-metre relative humidity ($r_{2\mathrm{m}}$), 2-metre temperature ($t_{2\mathrm{m}}$) and mean sea-level pressure ($msl$). The increments result from assimilating the full radar reflectivity field within the radar disc using the NN-based observation operator and the RF-based background-error covariance. The RF iterations are 4 and the correlation radius is 10\,Km. The black circle denotes the radar disc and the inner circle the location of the assimilated cluster.}
    \label{fig:increments_full}
\end{figure}

As in the previous section (\cref{subsec:blob_obs}), the localisation mask is set to 1 at all grid points inside the radar disc and to 0 elsewhere, ensuring that observation information is not propagated by the gradient of $\mathcal{H}$ beyond the radar disc. Nevertheless, the modelled auto-covariance through the RF in the $\mathbf{B}$-matrix spreads the observational impact slightly across the disc boundary (\cref{fig:increments_full}). Consequently, the resulting analysis increment pattern reflects the combined effect of the RF-based background autocovariances and the spatially extended, nonlinear mapping provided by the NN-based $\mathcal{H}$. Since the applied $\mathbf{B}$-matrix does not include vertical covariance components, increment signals at closely spaced vertical levels are not necessarily correlated (although vertical correlations can clearly be observed even without an explicit specification in $\mathbf{B}$, e.g. between temperature increments at 925 and 975 hPa, $u$ and $v$ increments at 925 and 850 hPa, or even across variables - compare $r2m$ and $t2m$ increments). The properties of the applied $\mathbf{B}$-matrix are further illustrated through the single-observation experiment described in \ref{appendix:single_obs}.

\begin{figure}
    \centering

    \begin{subfigure}[h]{0.99\textwidth}
        \includegraphics[width=\textwidth]{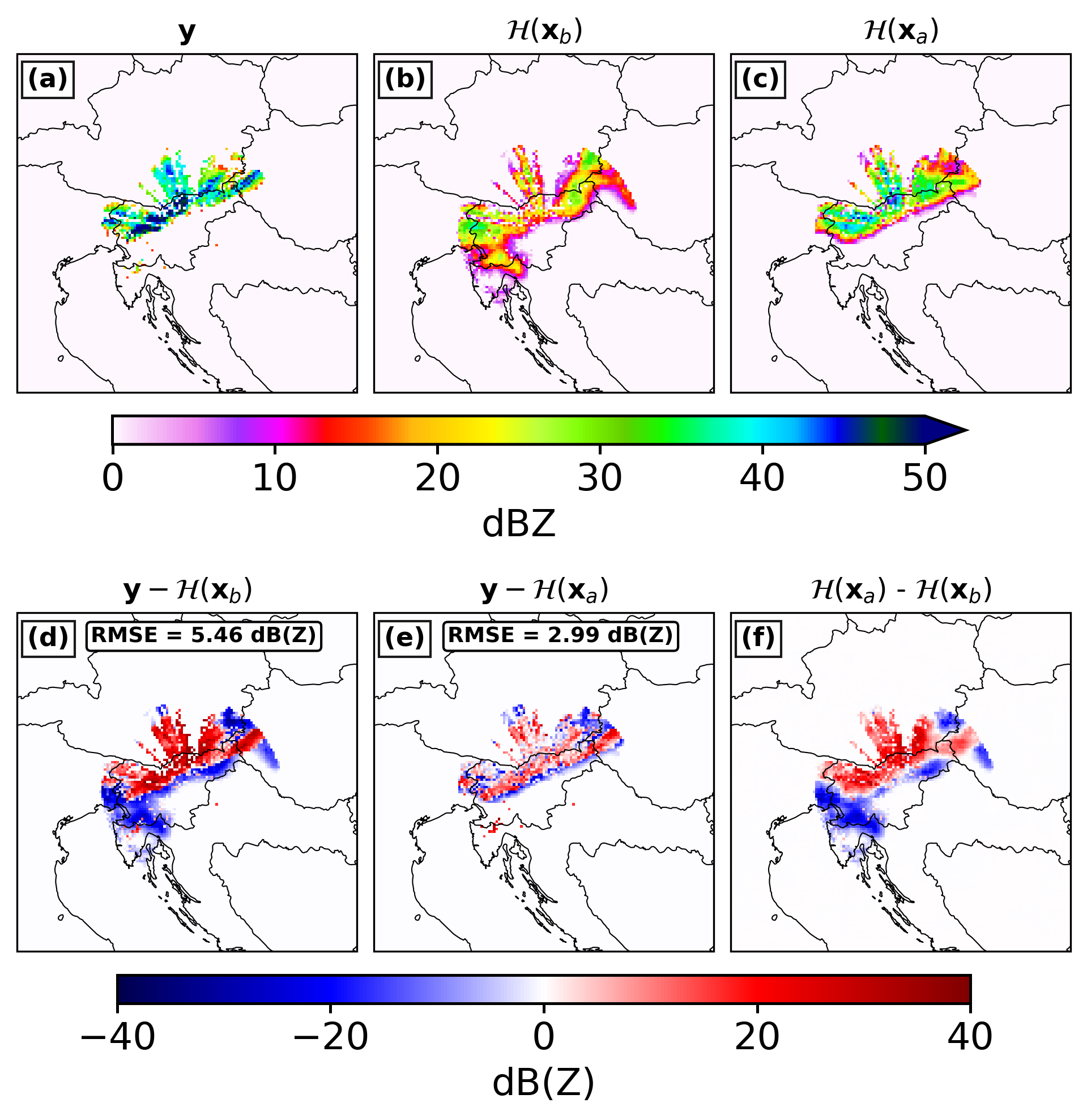}
    \end{subfigure}
    \caption{
    Evaluation of the NN-based observation operator for the Slovenian floods case on 4~August~2023 at 00~UTC. 
    Top row: (a) observed radar reflectivity, (b) background model-equivalent reflectivity $\mathcal{H}(\mathbf{x}_b)$, and (c) analysis model-equivalent reflectivity $\mathcal{H}(\mathbf{x}_a)$. Bottom row: (d) innovation $\mathrm{\mathbf{y}} - \mathcal{H}(\mathbf{x}_b)$ with RMSE $=5.46\,\mathrm{dB(Z)}$, (e) residual $\mathrm{\mathbf{y}} - \mathcal{H}(\mathbf{x}_a)$ with RMSE $=2.99\,\mathrm{dB(Z)}$, and (f) change in model-equivalent reflectivity $\mathcal{H}(\mathbf{x}_a)-\mathcal{H}(\mathbf{x}_b)$.}
    \label{fig:HxbHxa}
\end{figure}

The resulting multivariate increments are shown in Fig.~\ref{fig:increments_full}. The increments exhibit a mesoscale structure, with alternating positive and negative anomalies of temperature, wind, humidity, mean sea level pressure, and 2\,m variables that follow the organisation of the observed reflectivity system in \cref{fig:HxbHxa}a. Some increments, e.g. lower level $u,v,r$, exhibit localized dipolar or elongated features near regions of strong reflectivity gradients, suggesting that the NN-based observation operator can, in some cases, project information from the complex radar pattern onto dynamical and thermodynamical components of the model state. At the same time, the increments remain confined to the radar disc and decay smoothly towards its edge, demonstrating that the combination of the localisation mask and the RF autocovariance successfully limits the spatial extent of the impact of the observations. Nevertheless, the detailed relationships among the increments should not be given a fully physical interpretation, because the construction of the simplified $\mathbf{B}$-matrix does not permit such conclusions. The present experiment is therefore intended primarily as a proof of concept.

The impact of assimilating the full radar disc on the reflectivity field itself is summarised in \cref{fig:HxbHxa}. The top row shows: the observed reflectivity (\cref{fig:HxbHxa}a), the model equivalent reflectivity from the background $\mathcal{H}(\mathbf{x}_b)$ (\cref{fig:HxbHxa}b) and the model equivalent reflectivity from the analysis $\mathcal{H}(\mathbf{x}_a)$ (\cref{fig:HxbHxa}c). The analysis field clearly aligns more closely with the observation, particularly along the main high-reflectivity band and in the upstream convective cells. The precipitation pattern is translated towards the northwest in alignment with radar observations.

The bottom panels display the innovation $\mathrm{\mathbf{y}} - \mathcal{H}(\mathbf{x}_b)$ (\cref{fig:HxbHxa}d) and residual $\mathrm{\mathbf{y}} - \mathcal{H}(\mathbf{x}_a)$ (\cref{fig:HxbHxa}e), together with their domain-averaged RMSE. Finally, \cref{fig:HxbHxa}f, shows $\mathcal{H}(\mathbf{x}_a)-\mathcal{H}(\mathbf{x}_b)$. Assimilation of the radar data reduces the RMSE from 5.46\,dB(Z) to 2.99\,dB(Z), indicating a substantial improvement in the consistency between the model state and the observed reflectivity. Residual discrepancies are mainly confined to small-scale features and sharp gradients, which are only partially represented in the model fields and therefore cannot be fully corrected by the NN-based observation operator. Overall, the case study demonstrates that the proposed system can effectively pull the model state towards complex radar observations while preserving a physically reasonable multivariate increment structure.

\newpage
\section{Discussion and Conclusions}\label{section:conclusions}

In this study, we have developed and tested a neural-network–based observation operator that maps short-range ALADIN model forecasts of temperature, humidity, horizontal wind components, and mean sea-level pressure to weather radar reflectivity within a 3DVar data assimilation framework. Using five years of radar and model data, we trained an encoder–decoder ResUNet to emulate the nonlinear relationship between model fields and observed reflectivities. Statistical evaluation (\cref{subsec:stat}) showed that the trained observation operator
is able to represent the broader distribution and spatial occurrence of weak-to-moderate precipitation but remains limited in reproducing the frequency, intensity, and spatial organization
of the strongest radar reflectivity features. We then embedded this operator into a 3DVar system with a simplified background-error covariance model to assess the behaviour of the operator in controlled experiments. The control vector was restricted to dynamical and thermodynamical variables and did not include hydrometeors. This setup was intentionally designed to assess whether radar reflectivity can be used not only to correct variables directly linked to the observed signal, as is commonly done in existing approaches, but also to constrain variables whose impact on forecast skill may persist longer. Such a formulation provides a complementary perspective to conventional radar data assimilation strategies, in which hydrometeor adjustments often dominate but may remain comparatively short-lived. At the same time, hydrometeors are essential for representing the microphysical structure of convection, and their future inclusion in the control vector would likely improve the physical consistency of the analysed state by linking processes across multiple scales.

By learning a nonlinear mapping between dynamical and thermodynamical model states and observations, this data-driven observation operator offers a flexible, differentiable alternative to traditional parameterised simulators of radar reflectivity. Our formulation allows, for the first time, to update dynamic and thermodynamic variables in the model state from the observed radar reflectivity data. The single observation experiment, the experiment assimilating a compact blob of 250 radar pixels  (\cref{subsec:blob_obs}), and the August 2023 Slovenian floods case (\cref{subsec:slo_case}) all show that radar reflectivity innovations are translated into analysis increments that modify humidity, temperature, wind, and mean sea-level pressure, rather than merely adjusting the hydrometeor fields. The resulting analysis fields produce model-equivalent reflectivities that closely match the observed radar data. This focus on dynamical and thermodynamical variables was a deliberate choice in this proof-of-concept study. Rather than correcting the variables most directly linked to reflectivity, as is commonly done in existing approaches, we aimed to test whether radar observations can also constrain the dynamic variables, whose impact may be more persistent in convective-scale forecasts. At the same time, hydrometeors remain crucial for representing the microphysical structure of convection, and their future inclusion in the control vector will likely improve the physical consistency of the analysed state.

A key property of the NN-based operator is that, because of its convolutional architecture, each simulated radar pixel is influenced by a broader neighbourhood of model state variables, allowing $\mathcal{H}$ to exploit mesoscale patterns and gradients rather than purely local column information. However, most of the information is still gathered from the local proximity of the observation. As a consequence, its Jacobian spreads observational information along the horizontal structure of precipitation systems, and increments align with the observed precipitation bands. In other words, the operator not only projects model variables into reflectivity space but also implicitly encodes cross-variable and cross-level relationships, as suggested by the single-observation and observation cluster experiments.

Despite these promising results, the present implementation has several limitations. First, we have not yet evaluated the impact of the NN-based operator on subsequent model forecasts. Because all experiments end at the analysis stage, no conclusions can yet be drawn regarding their effects on forecasting skill. Second, in this proof-of-concept configuration, hydrometeors were not included as input variables for training the NN operator to simulate radar reflectivity. In many operational radar DA systems, hydrometeors or precipitation-related variables can be directly updated through the DA, and the lack of such variables in our control vector limits the analysis ability to directly adjust microphysical fields, even when reflectivity is well fitted. Third, a limited number of vertical levels was used for the construction of observation operator. Fourth, the applied background-error covariance model is univariate: $\mathbf{B}$ is built assuming diagonal variances and an isotropic recursive filter that represents only horizontal autocovariances. Vertical autocovariances and cross-covariances between different model variables are neglected, which limits the physical interpretation of the produced increments.

More fundamentally, the proposed NN-based observation operator offers both advantages and challenges compared with traditional physically based radar operators. On one hand, it avoids some explicit assumptions related to microphysics and electromagnetic scattering, and it is able to reproduce complex spatial structures present in the observed radar signal. On the other hand, because the operator is learned from observations, it may also incorporate features of the observing system or training dataset that are not causally represented in the model state, such as unresolved processes, sampling inconsistencies, representativeness errors, or radar-specific artifacts. Therefore, good agreement between $\mathcal{H}(\mathbf{x})$ and the observed reflectivity does not necessarily mean that the physical relationship is fully captured or that the model correctly represents precipitation dynamics.

Accordingly, the innovation $\mathbf{y} - \mathcal{H}(\mathbf{x}_b)$ is less directly interpretable than with a conventional physically based forward operator. In the latter case, the innovation can often be attributed more directly to deficiencies in the model state. In the present framework, the observation operator might learn to offset the forecast/model error. However, part of the mismatch may reflect limitations of the learned mapping itself, including statistical compensations acquired during training, in addition to actual meteorological differences between the model background and reality. The operator may also generalize poorly when applied to situations insufficiently represented in the training dataset or when model-error characteristics differ markedly from those seen during training.

Furthermore, while the analysis is utilised for training the NN as the best available estimate of the atmospheric state, it inherently contains errors derived from the assimilation procedure, the underlying observations and the background state. 
The reliance on analyses is justified, as no single observing system can provide a comprehensive, instantaneous description of the full atmospheric state across all chosen input variables. Consequently, discrepancies may arise where the model state is not conducive to precipitation formation, yet precipitation is observed by radar. Despite such inconsistencies, the analysis fields represent the most physically consistent approximation of reality and therefore utilised as the primary input for training the neural network.

For all these reasons, the present study should be regarded mainly as a proof of concept. It demonstrates that an NN-based observation operator for complex data can be embedded consistently within a variational DA framework and can generate structured multivariate increments, but its physical robustness and forecasting value still require further investigation and development.

Future work will therefore focus on addressing these limitations. A priority is to extend the control vector to include hydrometeor species and/or related microphysical variables, so that the NN-based operator can be used in combination with more physically explicit precipitation adjustments, as routinely done in operational systems. In parallel, we plan to replace the current univariate $\mathbf{B}$ with a more realistic multivariate background-error covariance model that captures cross-variable and cross-scale balances, for example, by exploiting latent-space DA approaches based on autoencoders and unified neural covariance models (e.g. \citealp{melinc20243d} and  \citealp{melinc2025unified}). Finally, we aim to conduct full-cycle experiments using short-range forecasts in a regional domain, multiple radars, and additional case studies to quantify the impact of this framework on thunderstorm predictability.

More broadly, our results illustrate both the potential and the difficulties of integrating machine learning into geophysical DA systems. On the positive side, NN-based observation operators can capture complex, nonlinear relationships that are difficult to express analytically, while remaining fully differentiable and thus compatible with variational methods. We hope that the methodology and open datasets presented here will help the DA community to more systematically evaluate NN-based observation operators, encourage transparent diagnostic practices, and accelerate the transition from proof-of-concept prototypes to robust, operationally relevant systems.

\section*{Funding}
This publication is supported by the European Union’s Horizon Europe research and innovation program under the Marie Sklodowska-Curie COFUND Postdoctoral Programme, co-funded under the grant agreement No.101081355 - SMASH and by the Republic of Slovenia and the European Union from the European Regional Development Fund.

This research was supported by the Slovenian Research And Innovation Agency (Javna agencija za znanstvenoraziskovalno in inovacijsko dejavnost RS) research core funding No. P1-0188. This research was further supported by the University of Ljubljana Grant SN-ZRD/22-27/0510. \v{Z}iga Zaplotnik acknowledges the funding by the European Union under the Destination Earth initiative and Copernicus Climate Change Service (C3S).

\section*{Author Contributions}
M.S.\,-\,Data curation, Formal analysis, Software, Validation, Visualisation, Writing (original draft preparation), Conceptualisation, Methodology, Investigation, Funding acquisition, Project administration, Writing (review and editing), \v{Z}.Z.\,-\, Conceptualisation, Methodology, Investigation, Funding acquisition, Project administration, Writing (review and editing), G.S.\,-\, Investigation, Funding acquisition, Project administration, Writing (review and editing).

\section*{Acknowledgements}

MS thanks the Theoretical and Scientific Data Science group at SISSA for hospitality during the first secondment. In particular, Roberto Trotta and Serafina Di Gioia for their valuable discussions and suggestions. MS also thanks the Slovenian Met Office (ARSO) for hospitality during the second secondment and for providing the ALADIN model and radar data used in this study. In particular, Benedikt Strajnar, Jure Cedilnik, and Neva Pristov for valuable discussions and suggestions. Finally, MS thanks Uro\v{s} Perkan for valuable discussions and suggestions.

\section*{Code and data availability}
All codes used in this study are written in Python and, together with the datasets, are publicly archived on Zenodo under the Creative Commons Attribution 4.0 International (CC BY 4.0) license.

The nonlinear neural-network-based observation operator used to map ALADIN numerical weather prediction model outputs to radar–reflectivity observation space is available as the repository “3DVar Neural Network-Based Observation Operator” under DOI: 10.5281/zenodo.17898083 \citep{Zenodo3DHNNmodel}. The directory structure includes a \texttt{TEST} subdirectory containing a ready-to-use trained ResUNet model. The dataset necessary to train the model is “LISCA-ALADIN HNN” under DOI:10.5281/zenodo.17880622 \citep{Zenododatasets}.

The 3DVar data assimilation system is also archived on Zenodo as the repository “3DVar for Neural Network-Based Observation Operator” under DOI: 10.5281/zenodo.17899024 \citep{Zenodo3DVar}. It contains the 3DVar configurations used in this study and auxiliary datasets (precomputed $\mathbf{B}$-matrix and the observation-space mask) needed to run and test the system.

\section*{Conflict of Interest Statement}
The authors declare no conflicts of interest.

\section*{Disclaimer}
Co-funded by the European Union. Views and opinions expressed are, however, those of the author(s) only and do not necessarily reflect those of the European Union or European Research Executive Agency. Neither the European Union nor the granting authority can be held responsible for them.

\clearpage
\bibliographystyle{Copernicus}
\bibliography{main.bib}

\appendix
\renewcommand{\thefigure}{A\arabic{figure}}
\setcounter{figure}{0}
\section*{Appendices}\label{appendices}
\addcontentsline{toc}{section}{Appendices}
\section{Neural Network Architecture}\label{appendices:A}

The observation operator $\mathcal{H}$ is estimated using a residual U-Net (ResUNet, \citealp{he2015deep}) to map multi-level meteorological model fields to radar reflectivity observations at the same spatial resolution. 
The input tensor $\mathbf{x} \in \mathbb{R}^{B \times 19 \times H \times W}$ comprises 19 channels of model fields: four pressure levels (975\,hPa, 925\,hPa, 850\,hPa and 800\,hPa ) of temperature ($t$), zonal wind ($u$), meridional wind ($v$), and relative humidity ($r$), comprising 16 channels, together with three surface variables: 2\,m temperature ($t{2m}$), 2\,m relative humidity ($r{2m}$) and mean sea-level pressure ($msl$). 
The model outputs a single-channel radar field $\hat{\mathbf{y}} \in \mathbb{R}^{B \times 1 \times H \times W}$. The input and output fields share the same spatial dimensions, both having $H=W=137$.

The architecture follows an encoder-decoder topology with symmetric skip connections between corresponding resolutions. 
Each encoder ($\mathbf{E}$) stage consists of a residual block:
\begin{equation}
  \begin{aligned}
  \mathbf{E}_{1} &= R_1(\mathbf{x})\\
  \mathbf{E}_{i} &= R_i(\mathrm{MP}_2(\mathbf{E}_{i-1})), \quad i = 2,\,3,\,4  
  \end{aligned}
\end{equation}

where $\mathrm{MP}_2$ denotes $2\times2$ max pooling, and $R_i$ represents a residual block composed of two $3\times3$ convolutions with Batch Normalisation and LeakyReLU activations, plus a skip connection with projection when the number of channels changes. 
Within each block, a Squeeze-and-Excitation (SE) unit \citep{hu2018squeeze} performs channel-wise attention by scaling feature maps according to a learned relevance weight derived from global average pooling. 
The encoder feature dimensions are $\{16, 32, 64, 128\}$.

The decoder ($\mathbf{D}$) performs progressive upsampling via transposed convolutions and skip concatenations:

\begin{equation}
  \begin{aligned}
\mathbf{D}_j
= R_j\!\left(
\mathrm{TC}_2(\mathbf{E}_j)
\;\oplus\;
C(\mathbf{E}_{j-1})
\right) \quad j = 4,\,3,\,2
  \end{aligned}
\end{equation}
where $\mathrm{TC}_2$ denotes a $2\times2$ transposed convolution with stride~2, 
$C(\cdot)$ is a center-cropping operator ensuring spatial alignment and the simbol $\oplus$ represents concatenation along the channel dimension. The decoder feature dimensions are $\{64, 32, 16\}$. 
A final $1\times1$ convolution layer projects the decoder output to one channel:
\begin{equation}
\hat{\mathbf{y}} = \mathrm{Conv}_{1\times1}(\mathbf{D}_2)
\end{equation}

where $\hat{\mathbf{y}}$ is the model equivalent reflectivity.
To preserve grid consistency, all outputs are adjusted using symmetric reflect padding and cropping operations so that $\mathrm{size}(\hat{\mathbf{y}}) = (H, W)$, even for odd input dimensions (e.g., $137\times137$).

The ResUNet thus combines multiscale feature extraction, residual learning, and channel attention in a compact architecture that effectively translates numerical weather prediction (NWP) predictor fields into spatially coherent radar observations.

\subsection{Loss Function}

To guide the model toward physically meaningful learning, we introduce a spatially and structurally
aware loss weighting strategy. The objective is not to alter the definition of the loss terms themselves,
but to control the spatial distribution of their influence during training.

\subsubsection{Gaussian Spatial Weighting Based on Radar Coverage}

To reflect the varying reliability of radar observations with distance from the radar location \citep{sebastianelli2013precipitation}, we define a spatial weighting field that decays smoothly as a Gaussian function.

After preprocessing, the radar and model share the same spatial grid, but valid observations are confined to a circular domain of a radius of 160~km (\cref{fig:grid}).
Since measurement uncertainty and beam broadening increase with distance, pixels farther from the radar are assigned lower weights in the loss computation.

The weighting map is first initialised by identifying the spatial extent of valid reflectivity data, producing a binary mask of observed regions, which is then combined with a Gaussian decay computed from the Euclidean distance, $d(x,y)$, to the radar centre. The Gaussian profile is defined as

\begin{equation}
W_g(x, y) =
\begin{cases}
w_\mathrm{g_{inner}}, & \text{if } d(x,y) < r_{\mathrm{inner}} \\[5pt]
\exp\!\left(-\frac{d(x,y)^2}{2\sigma^2}\right), & \text{otherwise}
\end{cases}
\label{eq:gaussian}
\end{equation}

\begin{figure}[t]
    \centering
    \includegraphics[width=0.65\linewidth]{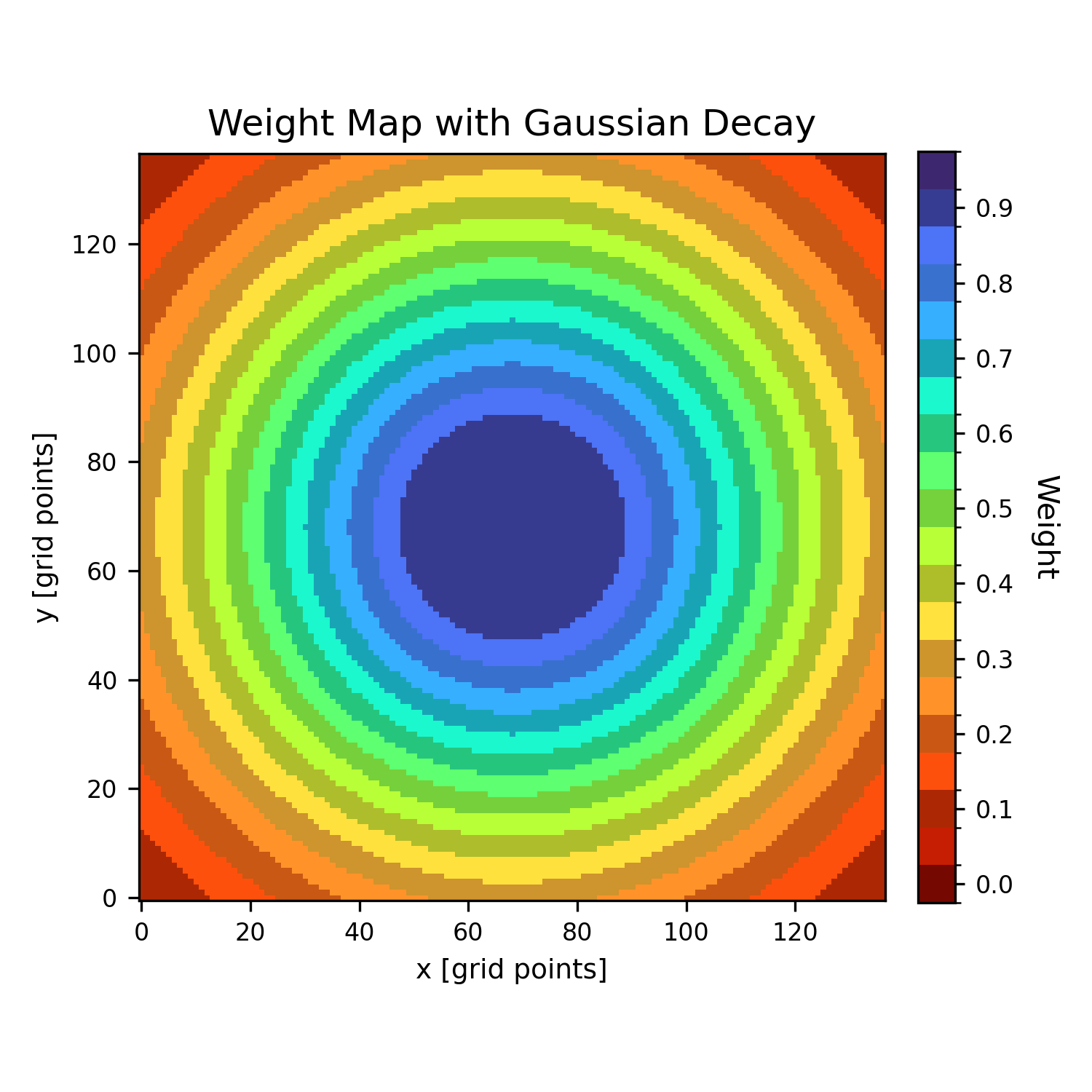}
    \caption{Spatial weighting map used in the loss function. 
    The weights follow a Gaussian decay centred at the radar location, 
    with higher importance near the radar and reduced weights farther away.}
    \label{fig:gaussian_weights}
\end{figure}

For this study, we used $\sigma = 45$ km to control the rate of decay.
Within a central radius of $r_{inner}=20$~km, the weights are kept constant ($w_{g_\mathrm{inner}}=0.9$) to avoid overemphasising a very small core region, ensuring a smooth transition from the inner plateau to the decaying outer field. The inner plateau ensures uniform weighting near the centre, while the Gaussian tail gradually downweights peripheral regions.\\
This spatial weighting acts as a physically informed prior:
errors near the radar, where observations are most accurate, contribute more strongly to the loss, while those in peripheral regions, where measurement uncertainty is higher, are gradually downweighted. The resulting Gaussian weight field is shown in Fig.~\ref{fig:gaussian_weights}, illustrating the smooth radial decay centred at the radar position.

\subsubsection{Boundary-Enhanced Target Weighting}

Radar reflectivity fields often exhibit sharp gradients at storm boundaries. In this context, “storm boundaries” refer to boundaries in the radar-observed reflectivity field, which do not always coincide with the true physical extent of the storm. In particular, for shallow systems (e.g. during the cold season or at long ranges), the radar beam may overshoot hydrometeors, so that the apparent echo edge in reflectivity is displaced relative to the actual precipitation boundary. Consequently, we introduce a boundary-emphasis map $W_b$ derived from the ground-truth field to preserve such discontinuities. $W_b$ emphasises transitions in the observation space defined by the radar product, rather than enforcing strict physical storm edges. This mechanism is designed to assign higher importance to pixels located along the transition between precipitating and non-precipitating regions, where small spatial displacements may produce substantial structural errors. Following, we discuss the methodology used to construct $W_b$.

\paragraph{Binary mask construction}
During the training, the NaN reflectivity values are set to 0. Let \(T \in \mathbb{R}^{H \times W}\) denote the target reflectivity field.  
We first construct a binary mask
\begin{equation}
M(x,y) = \begin{cases}
1, & T(x,y) > 0 \\
0, & T(x,y) = 0
\end{cases}
\end{equation}
which identifies the support of non-zero reflectivity.

\paragraph{Morphological dilation and erosion}
To extract the boundary of the support region, we apply dilation, $\mathrm{Dil}(\cdot)$, and erosion, $\mathrm{Ero}(\cdot)$, using a square
structuring element of width \(2r+1\), where \(r\) is the user-specified boundary radius in pixels (3 in this study). In the implementation, both operations are performed using two-dimensional
max-pooling layers:
\begin{equation}
\mathrm{Dil}(M) = \text{maxpool}(M;\, 2r+1, \text{stride}= \text{padding}=r)
\end{equation}
\begin{equation}
    \mathrm{Ero}(M) = 1 - \text{maxpool}
(1 - M;\, 2r+1, \text{stride}=1, \text{padding}=r)
\end{equation}

The dilation operation expands the support region outward by \(r\)~pixels, whereas erosion
shrinks it inward by the same amount. Their difference characterizes the thin ring enclosing
the original boundary.

\paragraph{Boundary-ring extraction}
The raw boundary mask is obtained as
\begin{equation}
R = \max\bigl( \mathrm{Dil}(M) - \mathrm{Ero}(M),\, 0 \bigr)
\end{equation}
which yields a binary ring of approximate thickness \(2r\) surrounding the transition zone
between zero and nonzero reflectivity. Pixels in this ring correspond to locations where edge
fidelity is most relevant for visual and physical accuracy.

\begin{figure}[t]
    \centering
    \includegraphics[width=0.65\linewidth]{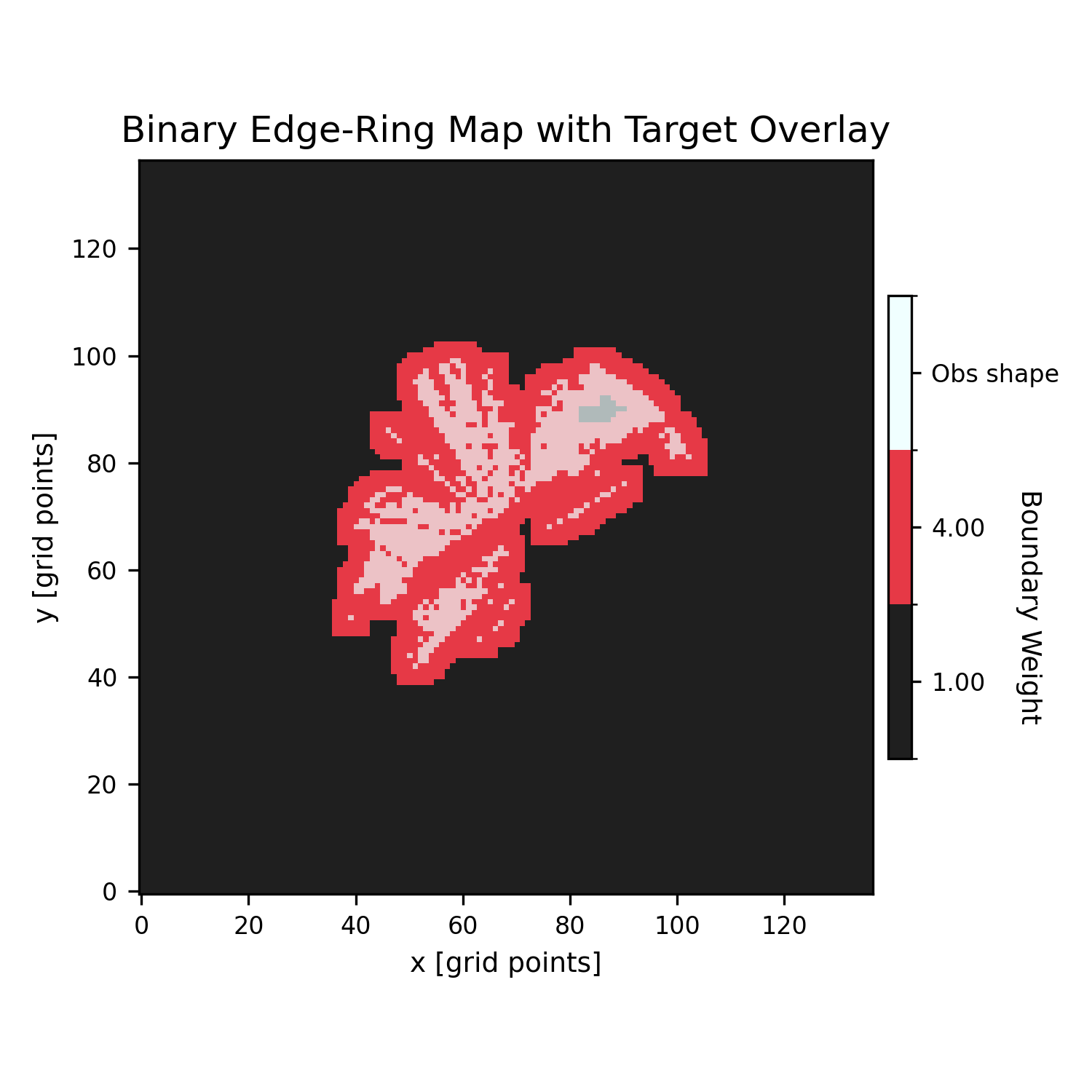}
    \caption{
Boundary-enhancement weighting map with overlaid observation footprint.
Red pixels indicate the boundary-weighting ring produced by the dilation--erosion
operation, where the loss is locally amplified to emphasise sharp reflectivity transitions.
Black pixels represent regions where no additional weighting is applied.
The light-blue overlay corresponds to the target observation region, illustrating the spatial extent of valid radar reflectivity data.
Together, these fields show how the boundary-weighting mechanism selectively increases the loss contribution only along the edges of the observed precipitation structures.
}
    \label{fig:boundary_weights}
\end{figure}

\paragraph{Weight assignment}
The final boundary-enhanced weight map is defined as
\begin{equation}
W_b(x,y) = 1 + \beta\, R(x,y)
\end{equation}
where \(\beta > 0\) is a tunable amplification factor (3 in this study). Pixels in the boundary ring thus receive
a weight of \(1+\beta\), while all other pixels retain unit weight.  
In practice, \(\beta\) values in the range \(2\)–\(4\) provide a moderate emphasis without destabilising optimisation. An example of boundary weights is shown in \cref{fig:boundary_weights}.

\paragraph{Effect on training}
This boundary emphasis increases the penalty for structural errors near storm edges, helping
to preserve sharp gradients, improve spatial coherence, and reduce the tendency of neural
networks to oversmooth precipitation boundaries.  
The method is computationally inexpensive, differentiable with respect to predictions, and
compatible with any pixelwise loss formulation.

\subsubsection{Combined Spatial Weight Map}

The total spatial weighting applied during training is obtained by the normalized product
\begin{equation}
W(x, y) = \frac{W_g(x, y)\, W_b(x, y)}{\overline{W_g W_b}}
\label{eq:combined}
\end{equation}
where the denominator denotes the mean over all spatial elements.
Normalization maintains a unit mean weight, thereby stabilizing 
the overall gradient magnitude and learning rate.

\subsubsection{Edge-Aware Weighted Loss}

Given a model prediction \( P \) and corresponding target field \( T \), the network is trained using the following composite loss function
\begin{equation}
\mathcal{L}
= \lambda_{\mathrm{Huber}}\,\mathcal{L}_{\mathrm{Huber}}
+ \lambda_{\mathrm{grad}}\,\mathcal{L}_{\mathrm{grad}}
+ \lambda_{\mathrm{SSIM}}\,\mathcal{L}_{\mathrm{SSIM}}
\end{equation}
where all pixelwise components are modulated by a shared spatially varying weight map $W$.
The first term, the Huber loss,
\begin{equation}
\mathcal{L}_{\mathrm{Huber}} = 
\frac{1}{N}\sum_{x,y} W(x,y)\, h\!\big(P(x,y) - T(x,y)\big)
\end{equation}
uses the Huber function \( h(\cdot) \)
\begin{equation}
h(x) =
\begin{cases}
\frac{1}{2}x^{2}/\delta, & |x| \le \delta \\[4pt]
|x| - \frac{\delta}{2}, & |x| > \delta
\end{cases}
\end{equation}
and provides robustness to outliers while preserving sensitivity to small errors.\\
Edge preservation is encouraged through the gradient-difference term
\begin{equation}
\mathcal{L}_{\mathrm{grad}} =
\frac{1}{N}\sum_{x,y} W(x,y)\, 
\big|\nabla P(x,y) - \nabla T(x,y)\big|
\end{equation}
where $\nabla(\cdot)$ denotes the Sobel operator. It is a gradient-based discrete differentiation filter that estimates local intensity variations for edge detection. In our implementation, we first apply reflect padding to the input image and then convolve it with two $3 \times 3$ Sobel kernels to obtain $x$ and $y$ derivative approximations.\\
Perceptual consistency is promoted through the Structural Similarity Index Measure (SSIM) loss
\begin{equation}
\mathcal{L}_{\mathrm{SSIM}} =
\frac{1}{N}\sum_{x,y} W(x,y)\,
\big[1 - \mathrm{SSIM}\big(P,T\big)\big]
\end{equation}

The multiplicative structure of Eq.~\eqref{eq:combined} 
ensures that the Gaussian prior \( W_g \) and boundary emphasis \( W_b \)
jointly modulate the contribution of each pixel to all loss terms. Regions near the radar centre or containing strong reflectivity receive proportionally larger gradients, while uniform or peripheral regions are downweighted.
This spatially adaptive weighting scheme effectively directs learning toward the
most informative regions of the radar field while maintaining stable optimisation dynamics.
Empirically, it enhances the preservation of fine-scale boundaries and high-intensity features, resulting in sharper and more physically coherent reflectivity reconstructions.

\section{The structure of $\mathbf{B}$-matrix}\label{appendix:single_obs}

This single-observation experiment is designed to isolate and interpret the local response of the 3DVar system to an individual radar reflectivity observation $y_{\mathrm{obs}}$.
This configuration provides a clean diagnostic of how the background-error covariance matrix $\mathbf{B}$, combined with the RF that models the horizontal auto-covariance, spreads the information of a pixel-wise increment in the surrounding area. 

The 3DVar problem (\cref{eq:J}) is then reformulated as follows:
\begin{equation}
\mathcal{J}(\mathbf{x}) =
\frac{1}{2}(\mathbf{x}-\mathbf{x}_{\mathrm{b}})^{\mathrm{T}}\mathbf{B}^{-1}(\mathbf{x}-\mathbf{x}_{\mathrm{b}}) + \frac{\left(y_{\mathrm{obs}}-\mathbf{m}^{\mathrm{T}}\mathcal{H}(\mathbf{x})\right)^2}{2\sigma_o^2} \, 
\label{eq:J_singleobs}
\end{equation}
where $y_{\mathrm{obs}}$ is the single observation, $\sigma_o$ its error standard deviation, $\mathbf{m}^{\mathrm{T}}$ is a row vector containing zeros and a single element with value of 1 representing the observed point. This ensures that the scalar modelled reflectivity value is compared to a single observation. 
Additionally, we reformulate the gradient of the cost function as
\begin{equation}
\nabla \mathbf{\mathcal{J}(x)} = \mathbf{B}^{-1}(\mathbf{x}-\mathbf{x}_{\mathrm{b}}) - \mathbf{M}_\mathrm{loc} \left( \frac{\partial \left(\mathbf{m}^{\mathrm{T}} \mathbf{\mathcal{H}(x)}\right)}{\partial \mathbf{x}} \right)^T \frac{y_{\mathrm{obs}} - \mathbf{m}^{\mathrm{T}}\mathbf{\mathcal{H}(x)}}{\sigma_o^2} \, 
\label{eq:gJ_single}
\end{equation}
where $\mathbf{M}_\mathrm{loc}$ is a binary localization mask matrix, which ensures that the observational part of the gradient only affects the state vector $\mathbf{x}$ variables in grid point collocated with the observation point.

This masking strategy serves as an explicit localisation: it prevents the convolutional structure of the NN-based observation operator from propagating the effect of a single radar pixel across the entire radar disc (defined by the black circle in \cref{fig:singleobs_increments}).

Fig. \ref{fig:singleobs_increments} shows the resulting 975 hPa temperature analysis increment following an assimilation of a single reflectivity observation extracted from the observed field at 00 UTC on 5 August 2023. The observation operator sensitivities are localised here to a grid-point of the observation, so that the increment only demonstrates the applied background-error covariance model. Increments in other fields (not shown) depict the same correlation structure which confirms the correctness of implementation. 

\begin{figure}[htbp]
     \centering
        \includegraphics[width=0.5\textwidth]{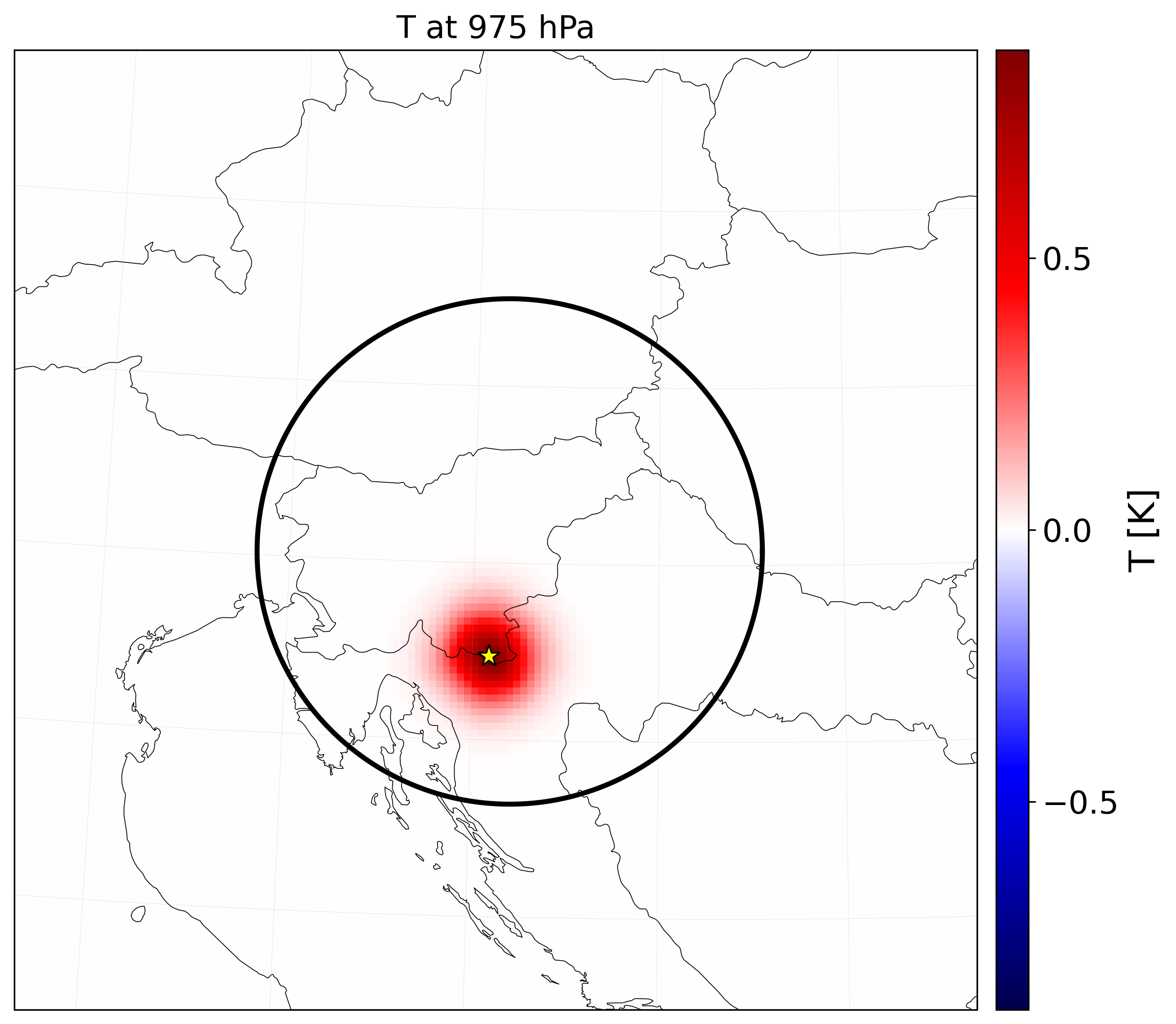}
    \caption{ A 3D-Var analysis increment in temperature at 975 hPa following the assimilation of a single reflectivity observation for the experiment initialized at 00 UTC on 5 August 2023. The yellow star marks the observation location, while the black circle delineates the radar-disc region. The red shading illustrates how the modeled horizontal autocovariance, represented by the RF, spreads the information from the initial temperature increment into the surrounding region. RF iterations are 4 and the correlation radius is 10\,km.}
    \label{fig:singleobs_increments}
\end{figure}

Horizontally, the increment field exhibits a compact, isotropic structure centred on the observation location, with amplitudes that decay smoothly toward zero with increasing distance, consistent with the prescribed autocorrelation structure of the background-error covariance. The first-order RF imposes a Gaussian-like horizontal autocorrelation, causing the information from the single observation to spread over a spatial scale of approximately 10\,km, in alignment with the prescribed correlation radius.



\subsection{Single observation assimilation}\label{subappendix:single_obs_2}

To further investigate the structure of the increments generated by the single-observation assimilation, we performed three different assimilation experiments without applying the localization mask. The analysis increments for three different observation locations, indicated by the stars are shown in \cref{fig:single_obs_increments_comparison}. The purpose of this experiment is to examine how the assimilation system distributes the information from a single reflectivity observation through the multivariate state space.

The three single-observation experiments show that the resulting increments remain mainly confined to the region surrounding the assimilated observation. Although the RF introduce a spatial spreading of the information, the amplitude of the increments rapidly decreases away from the observation location. This behaviour is consistent across the three selected points, indicating that the assimilation response is primarily controlled by the local influence of the observation, while the multivariate and spatial correlations provided by $\mathbf{B}$ and the RF determine the detailed structure of the surrounding increments.

\begin{figure}[p]
    \ContinuedFloat
    \centering

    \begin{subfigure}{1.0\textwidth}
        \centering
        \includegraphics[width=\linewidth]{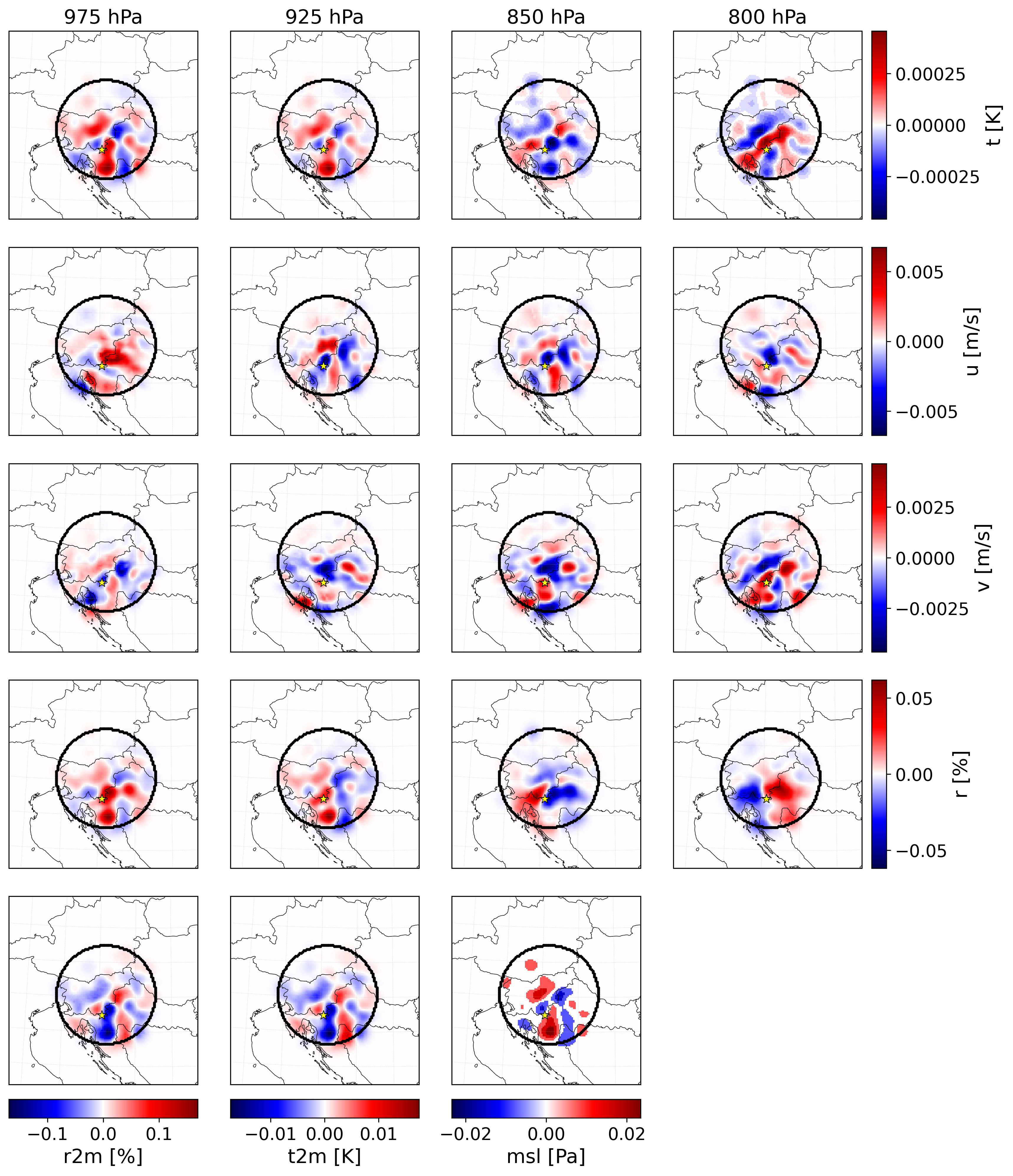}
        \caption{}
        \label{fig:single_obs_67_50}
    \end{subfigure}
\end{figure}

\clearpage

\begin{figure}[p]
    \centering
    \phantomcaption
    \label{fig:single_obs_increments_comparison}

    \begin{subfigure}{1.0\textwidth}
        \centering
        \includegraphics[width=\linewidth]{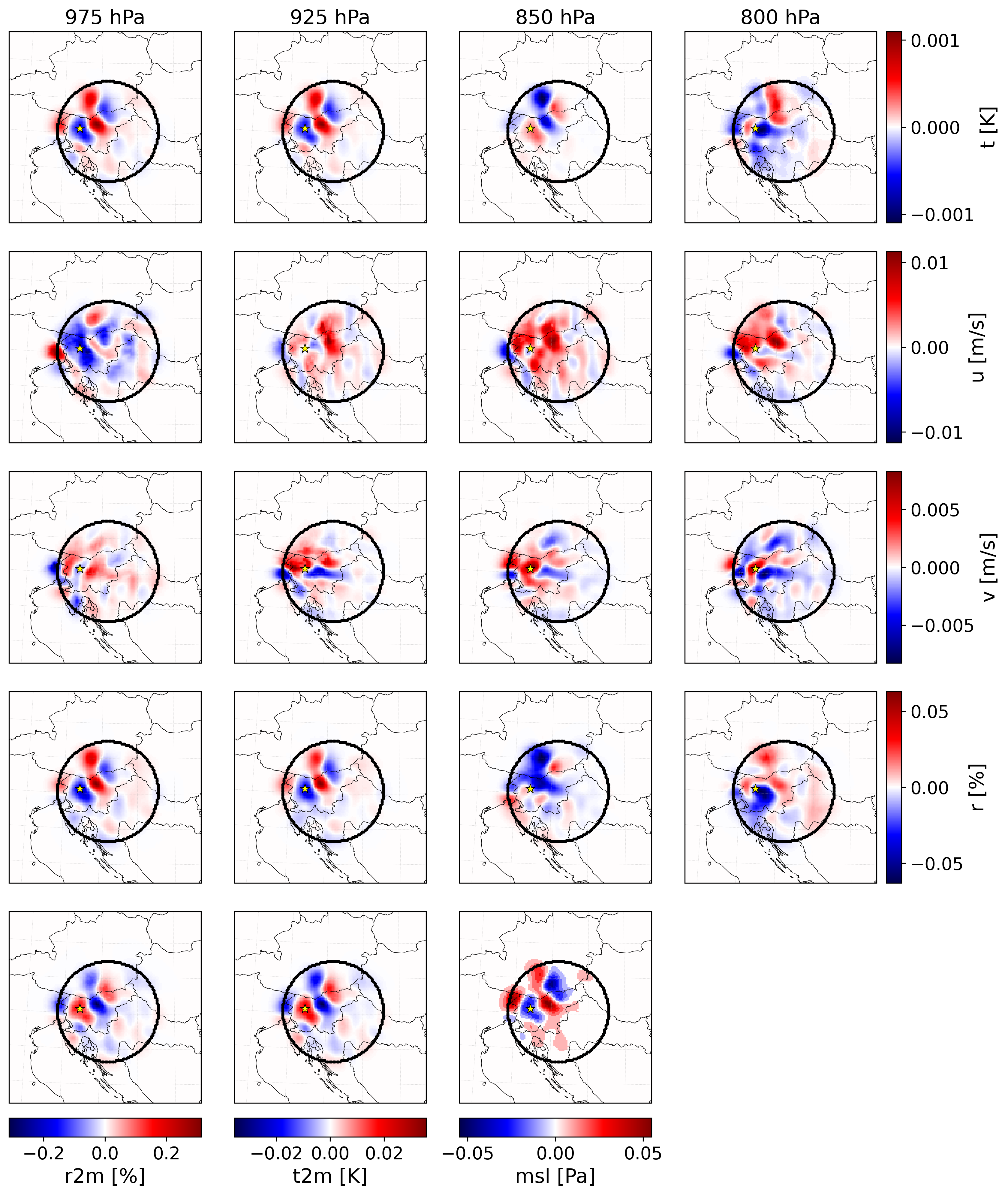}
        \caption{}
        \label{fig:single_obs_50_67}
    \end{subfigure}
\end{figure}

\clearpage

\begin{figure}[p]
    \ContinuedFloat
    \centering

    \begin{subfigure}{1.0\textwidth}
        \centering
        \includegraphics[width=\linewidth]{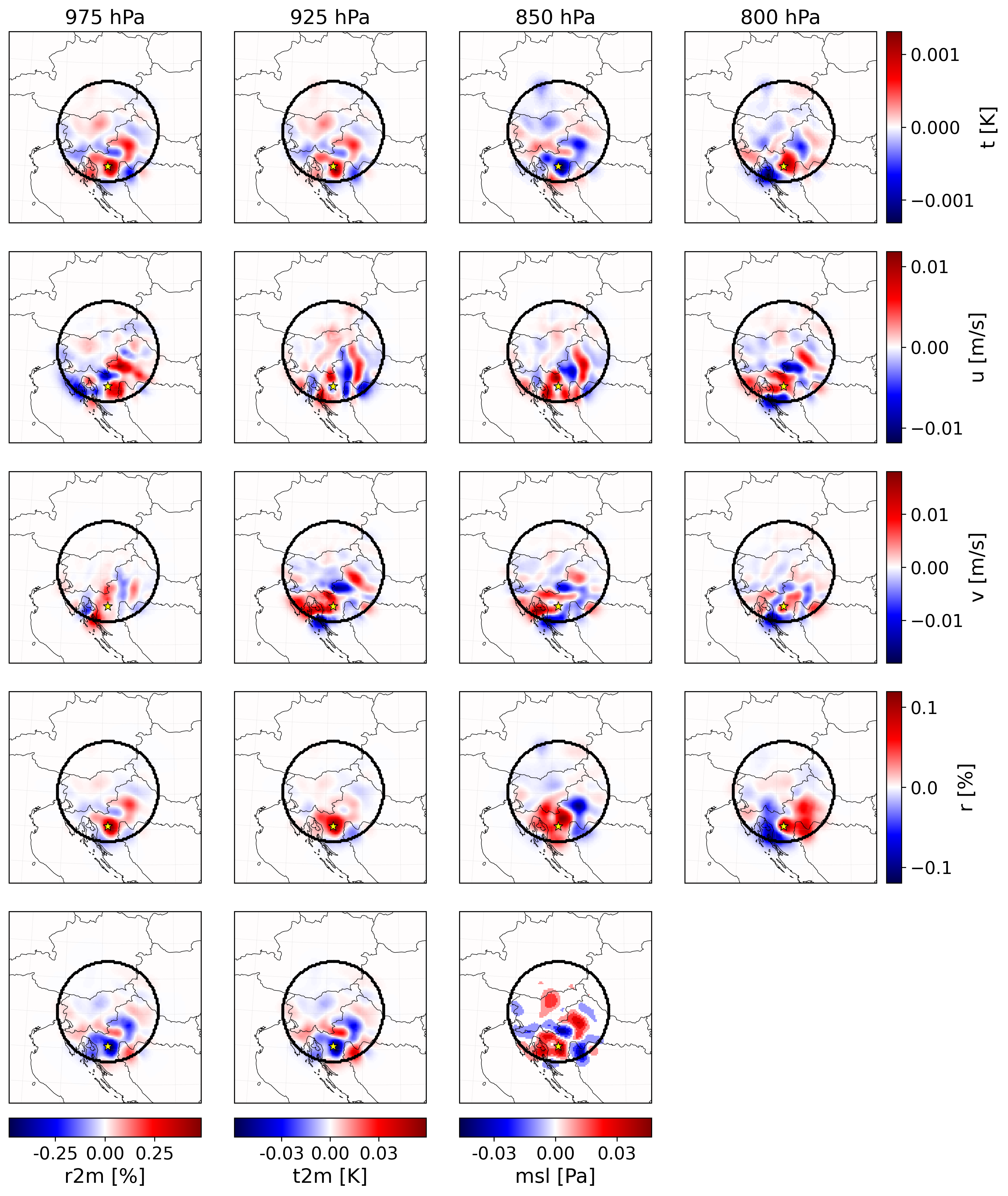}
        \caption{}
        \label{fig:single_obs_70_40}
    \end{subfigure}

    \caption{
    Single-observation assimilation experiments without localization mask. The analysis increments are shown for three different observation locations (marked by the stars) for the experiment of 5~August~2023 at 00~UTC. The experiments illustrate how the information from an isolated reflectivity observation is propagated through the multivariate background-error covariance matrix and the RF. Although the increments extend beyond the observation point, their amplitude remains mainly confined to the surrounding assimilation region, with a rapid decrease away from the observation location.
    }
    \label{fig:single_obs_increments_comparison}
\end{figure}


\end{document}